\documentclass[reprint,amsmath,amssymb,aps]{revtex4-2}

\usepackage{amsmath,amssymb,booktabs,braket}
\usepackage{graphicx}
\usepackage{dcolumn}
\usepackage{bm}
\usepackage{footnote}
\usepackage{hyperref}
\hypersetup{colorlinks=true, citecolor=blue, urlcolor=blue, linkcolor=blue}
\newcommand{\Tr}{\operatorname{Tr}}
\newcommand{\ii}{\mathrm{i}}

\begin{document}

\title{Polariton Bell Node for Quantum Repeaters}
\author{Junhui Cao$^{1}$}
\author{Alexey Kavokin$^{1,2,3,4}$}
\email{a.kavokin@westlake.edu.cn}
\affiliation{$^{1}$Abrikosov Center for Theoretical Physics, Moscow Center for Advanced Studies, Moscow 141701, Russia\\
$^{2}$School of Science, Westlake University, 18 Shilongshan Road, Hangzhou 310024, Zhejiang Province, China\\
$^{3}$Department of Physics, St. Petersburg State University, University Embankment, 7/9, St. Petersburg, 199034, Russia\\
$^{4}$Russian Quantum Center, Skolkovo, Moscow Region, 121205, Russia
}

\begin{abstract}
We propose a Bell-measurement node for quantum repeaters based on a planar semiconductor microcavity operating in the strong-coupling regime. Cavity photons hybridize with quantum-well excitons to form polaritons combining properties of photons and matter quasiparticles. A control photon loaded into one polariton mode changes the polarization response seen by a subsequently incident target photon. This conditional rotation is governed by the interplay of self-induced Larmor precession triggered by spin-dependent exciton-exciton interactions and the polarization beats caused by the splitting of transverse-electric and transverse-magnetic cavity modes. We identify conditions of the experiment that enable implementation of a controlled-Z gate and allow to distinguish all four Bell states in the ideal limit. The one-sided scattering scheme provides a lower interaction threshold than a scalar Kerr reference under the same assumptions. At a selected operating point, the bandwidth-induced identification error scales as the fourth power of the pulse bandwidth in the narrow-band limit. An additional fixed input rotation reduces this error even further. We describe the entanglement swapping between remote memories and determine the minimum quality of the elementary links needed to obtain an entangled output.
\end{abstract}
\maketitle

\paragraph*{Introduction.---}
Entanglement swapping connects neighboring elementary links through a Bell measurement on their inner qubits \cite{Zukowski1993,Briegel1998}. Photon loss, memory decoherence, and imperfect local operations limit the quantum repeater performance, which creates a major obstacle for the development of quantum information networks\cite{sangouard2011quantum,azuma2023quantum,lei2023quantum,hartmann2007role}. For polarization qubits, passive linear optics cannot deterministically identify all four Bell states without additional resources \cite{Lutkenhaus1999}. Nonlinear interactions can supply the required conditional operation, provided the logical branches emerge in sufficiently well-matched spectral modes. This requirement is particularly restrictive for traveling-wave optical nonlinearities \cite{Shapiro2006}. State-dependent cavity reflection offers an alternative way to control photonic phases \cite{DuanKimble2004}.

In this context, light-matter quasiparticles, exciton-polaritons, offer a valuable alternative to the repeaters based on absorption and emission of photons in quantum memory devices \cite{kavokin2022polariton,kavokin2017microcavities,10.1117/1.AP.8.5.056008,jia2026femtosecond}. Indeed, vacuum photons may convert to exciton-polaritons without any loss by coherent tunneling through the cavity mirrors. Moreover, spin-dependent polariton-polariton interactions provide a microscopic mechanism for conditional polarization rotation \cite{kavokin2003polarization,ryzhov2020spin,cao2026emergent}. Their anisotropy produces the effective field associated with self-induced Larmor precession \cite{Solnyshkov2007,krizhanovskii2006rotation,read2009stochastic,Sigurdsson2022}. Earlier polariton gate proposals use nonlinear interactions and pseudospin rotations in condensate or continuous-variable descriptions \cite{Solnyshkov2015,Kyriienko2016,Ghosh2020,Ricco2024}. A discrete two-polariton CZ gate has also been proposed using a free-electron ancilla \cite{Karnieli2024}. Experiments have demonstrated a single-polariton state entangled with an external photon \cite{cuevas2018first} and few-photon phase rotation \cite{Kuriakose2022}.

Here we propose an architecture of a polariton Bell node for the use in quantum quantum repeaters. Specifically, we consider a control excitation prepared before the target photon arrives. Its two spin amplitudes coherently select different target rotations, which form an entangling gate at suitable detunings and interaction strengths. We calculate the results of complete Bell measurement for finite pulses and demonstrate the entanglement being swapped onto the two remote memories.

\begin{figure*}[t]
\centering
\includegraphics[width=.9\textwidth]{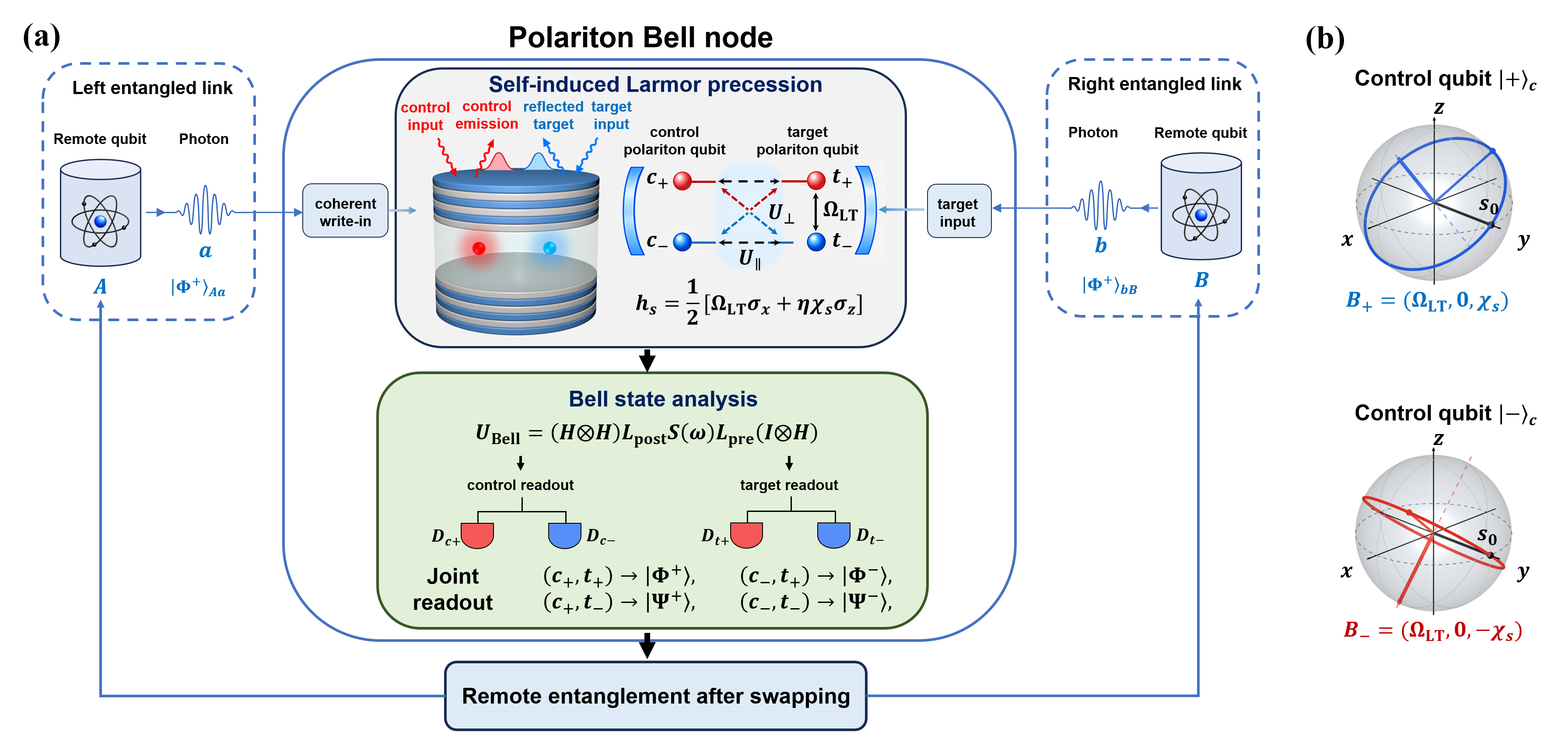}
\caption{Polariton Bell node and conditional spin rotation. (a) Conditional scattering and the local rotations in Eq.~(\ref{eq:full-analyzer}) map $(\Phi^+,\Phi^-,\Psi^+,\Psi^-)$ to outcomes $(00,10,01,11)$ at the chosen frequency $\omega_0$. Operators act from right to left. The optional input rotation $Q_\theta$ adjusts the measurement for a finite pulse. Each outcome specifies a Pauli correction on the remote pair. Loading and optical readout are assumed coherent interfaces. (b) The control states $|+\rangle_c$ and $|-\rangle_c$ select target fields $\bm B_\pm=(\Omega_{\rm LT},0,\pm\chi_s)$ in angular-frequency units. The condition $\Tr(R_+^\dagger R_-)=0$ makes the scattering gate locally equivalent to CZ.}
\label{fig:node}
\end{figure*}

\paragraph*{Microcavity modes and optical excitation.---}
We consider a planar semiconductor microcavity containing quantum wells at the antinodes of the confined photon mode. In the strong-coupling regime, the cavity photon and exciton resonances hybridize to form upper and lower polariton branches that feature an avoided crossing \cite{carusotto2013quantum}. The photon component couples these modes to external light, while the excitonic interactions detemine their nonlinearity \cite{ciuti1998role,glazov2009polariton}. The proposed node uses two distinguishable, optically addressable lower-polariton modes with overlapping spatial profiles. Each mode supports the circular-polarization states $|+\rangle$ and $|-\rangle$. We assume that the control mode can retain one excitation during scattering from the target mode.

A control photon is loaded first, preserving its polarization superposition. A target photon then probes the selected resonance and leaves through the reflected output, whose polarization depends on the control spin. After optical readout of the control, fixed polarization rotations and joint single-photon detection complete the measurement. The calculation assumes coherent loading and readout, with a single input--output port for the target. The splitting of transverse-electric and transverse-magnetic (TE--TM) optical modes contributes an in-plane effective field through the photonic component of the polariton \cite{glazov2005anisotropic,carusotto2013quantum}. We denote the resulting splitting of the selected target mode, in angular-frequency units, by $\Omega_{\rm LT}$ and choose its linear-polarization axes so that the field points along $x$.

\paragraph*{Conditional spin rotation.---}
Denote the selected control and target modes by $c$ and $t$. In the closed sector with one excitation in each mode, the spin-dependent interaction is
\begin{align}
H_{\rm int}={}&U_\parallel(n_{c+}n_{t+}+n_{c-}n_{t-})
+U_\perp(n_{c+}n_{t-}+n_{c-}n_{t+}) \notag\\
={}&U_0 I+\frac{\hbar\chi_s}{2}Z_cZ_t,
\quad \hbar\chi_s=U_\parallel-U_\perp .
\label{eq:ising}
\end{align}
Here $n_{m\pm}$ counts excitations of circular polarization $\pm$ in mode $m$, $Z_m$ is the corresponding Pauli operator, and $U_0=(U_\parallel+U_\perp)/2$. The energies $U_\parallel$ and $U_\perp$ are the interaction shifts of one polariton in each selected mode for equal and opposite circular polarizations. They include the excitonic fractions and spatial overlap of those modes \cite{glazov2009polariton}. The two spin channels involve different exchange and correlation processes. Experiments on GaAs-based structures show that the antiparallel-spin interaction can be attractive and is enhanced near a biexciton resonance, where dissipative nonlinearities also become important \cite{Takemura2017}. We use real effective interaction shifts in the coherent model and treat loss separately in the Supplemental Material.

For a fixed control spin, the interaction rotates a linearly polarized target about $+z$ or $-z$. A coherent control superposition selects both rotations at the amplitude level. Replacing the control spin by its mean would discard the correlations responsible for entanglement. For an interaction time $T$, the closed-system evolution is locally equivalent to CZ when $\chi_sT=\pi/2$ modulo $\pi$. At $\chi_sT=\pi$, it is the local product $-Z_cZ_t$ up to an overall phase. The control spin is assumed conserved during the interaction. Terms such as $c_+^\dagger c_-t_-^\dagger t_++{\rm h.c.}$ and transverse precession of the control are excluded from the effective model.

\begin{figure}[t]
\centering
\includegraphics[width=.9\columnwidth]{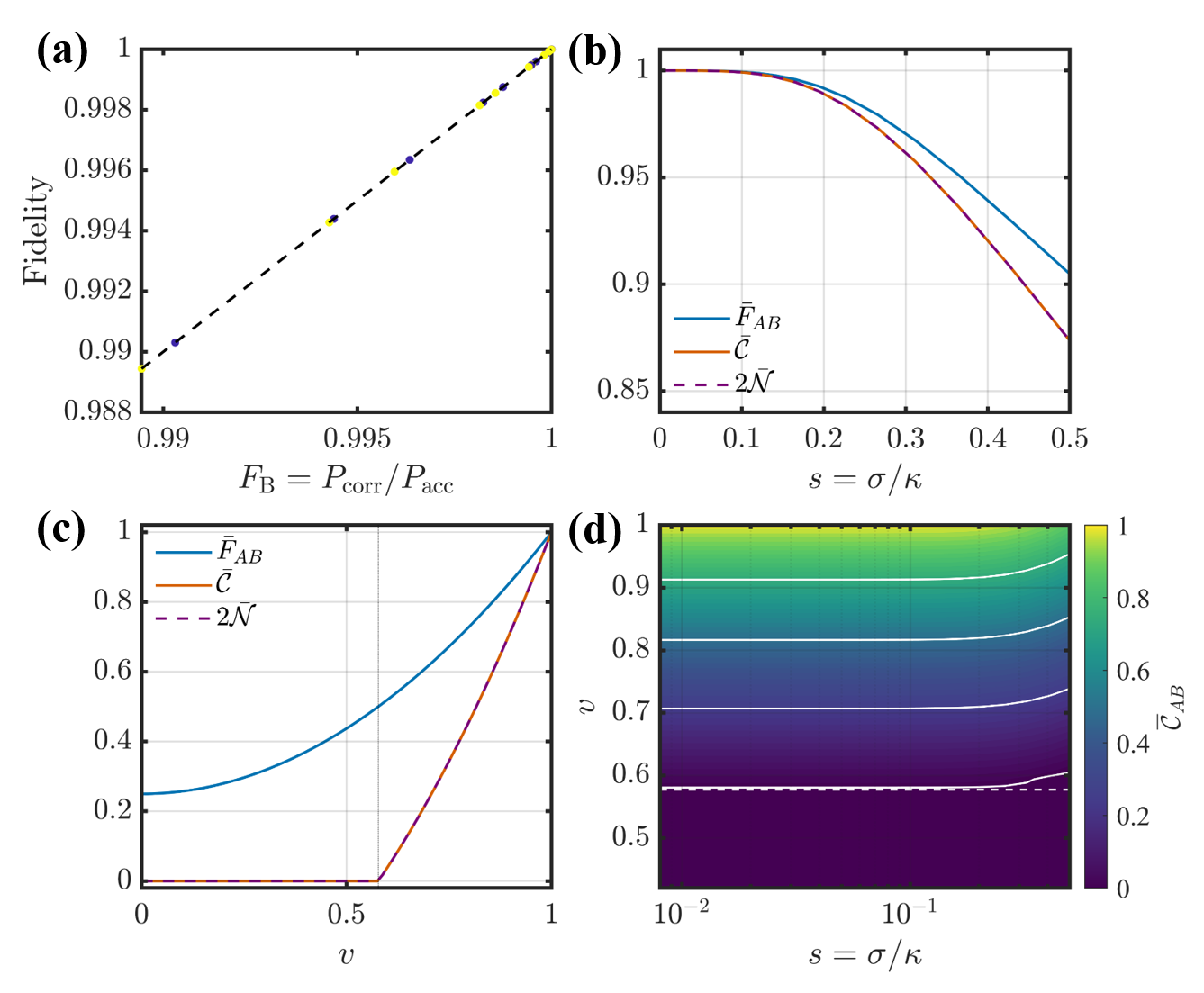}
\caption{Entanglement swapping with imperfect elementary links. (a) Outcome-averaged remote fidelity and $F_{\rm B}=P_{\rm corr}/P_{\rm acc}$ for ideal links, including sampled trace-nonincreasing operations. The dashed line is $\overline F_{AB}=F_{\rm B}$. (b) Remote Bell fidelity $\overline F_{AB}$, concurrence $\overline{\mathcal C}_{AB}$, and twice the negativity $2\overline{\mathcal N}_{AB}$ versus $s=\sigma/\kappa$ for ideal links. (c) Two identical Werner links and an ideal analyzer ($s=0$), giving remote visibility $v^2$ and entanglement for $v>1/\sqrt3$. (d) Mean remote concurrence from the full frequency integral and the two link-noise channels. The solid line is the threshold visibility $v_c(s)$ and the dashed line is $1/\sqrt3$. Panels (b) and (d) use $c=u=1/\sqrt2$, $x=0$, a lossless response, and the reference analyzer at $\gamma=0$. The narrow-band threshold is Eq.~(\ref{eq:remote-boundary}) at $X=0$. Averages retain the accepted outcome and its Pauli correction.}
\label{fig:swapping}
\end{figure}

Local phase rotations convert the entangling Ising evolution into CZ, and Hadamard rotations complete the Bell-state analyzer. For two initial links,
\begin{equation}
|\Phi^+\rangle_{Ac}|\Phi^+\rangle_{tB}
=\frac12\sum_{\mu=0}^{3}|B_\mu\rangle_{ct}
(I\otimes\sigma_\mu)|\Phi^+\rangle_{AB},
\label{eq:swap}
\end{equation}
up to the fixed Bell-state phase convention, with $\sigma_\mu$ a Pauli operator. The outcome $\mu$ determines a Pauli correction on the remote pair. For finite bandwidth and loss, let $E_\mu$ be the element of the positive-operator-valued measure (POVM) associated with an accepted outcome after tracing over frequency. Two ideal links give the unnormalized remote state
\begin{equation}
\widetilde\rho_{AB}^{(\mu)}=E_\mu^T/4,\qquad
p_\mu=\Tr E_\mu/4 .
\label{eq:choi-swap}
\end{equation}
The transpose is taken in the computational basis, before applying the Pauli correction. For $p_\mu>0$, the normalized state is $\rho_{AB}^{(\mu)}=e_\mu^T$, where $e_\mu=E_\mu/\Tr E_\mu$. Its probability-weighted fidelity after the corrections obeys $\overline F_{AB}=P_{\rm corr}/P_{\rm acc}=F_{\rm B}$. Here $P_{\rm corr}$ and $P_{\rm acc}$ are the correct-identification and acceptance probabilities for four equally likely Bell inputs. The ratio $F_{\rm B}$ is the probability of correct identification conditioned on acceptance. This relation, including loss, provides the starting point for the imperfect-link results in Fig.~\ref{fig:swapping}.

\paragraph*{Conditional scattering and the CZ gate implementation.---}
During target scattering, the control occupation remains $n_c=1$, and the target mode accesses its zero- and one-excitation sectors. In a lossless one-sided mode of linewidth $\kappa$, the response conditioned on the control spin $\eta=\pm1$ is
\begin{align}
h_\eta&=\frac12(\Omega_{\rm LT}\sigma_x+\eta\chi_s\sigma_z),\notag\\
R_\eta&=I-\kappa\left[\frac\kappa2I-\ii(\widetilde\Delta I-h_\eta)\right]^{-1}.
\label{eq:blocks}
\end{align}
Here $h_\eta=H_\eta/\hbar$ acts on the target polarization, and $\Delta$ is measured from its noninteracting resonance. The mean interaction shift gives $\widetilde\Delta=\Delta-U_0/\hbar$. Conservation of the control spin, $[H_{\rm node},Z_c]=0$, gives
\begin{equation}
S=|+\rangle\langle+|\otimes R_++|-\rangle\langle-|\otimes R_-.
\label{eq:controlled}
\end{equation}
At the pulse's central frequency $\omega_0$, diagonalize the relative unitary as $V_0=R_+^\dagger(\omega_0)R_-(\omega_0)=W_0\operatorname{diag}(e^{\ii\phi_0},e^{\ii\phi_1})W_0^\dagger$. The local rotations
\begin{equation}
\begin{aligned}
L_{\rm pre}&=I\otimes W_0,\\
L_{\rm post}&=\operatorname{diag}(1,e^{-\ii\phi_0})\otimes
[W_0^\dagger R_+^\dagger(\omega_0)],
\end{aligned}
\label{eq:local-factors}
\end{equation}
give $L_{\rm post}S(\omega_0)L_{\rm pre}=\operatorname{diag}(1,1,1,e^{\ii(\phi_1-\phi_0)})$. The eigenphase difference sets the controlled phase. With the Hadamard rotations, the reference analyzer is
\begin{equation}
U_{\rm car}(\omega)=(H\otimes H)L_{\rm post}S(\omega) L_{\rm pre}(I\otimes H),
\label{eq:full-analyzer}
\end{equation}
where the rightmost operation acts first and all local rotations remain fixed as $\omega$ varies. Equal determinants of $R_+$ and $R_-$ make $q=\Tr V/2$ real. At $q=0$, the eigenvalues of $V$ are opposite, and Eq.~(\ref{eq:local-factors}) converts the scattering operator into CZ. The reference analyzer then maps $(\Phi^+,\Phi^-,\Psi^+,\Psi^-)$ to $(00,10,01,11)$ at $\omega_0$, up to outcome phases. The eigenvalue ordering and eigenvector phase convention are given in the Supplemental Material.

Define $x=\widetilde\Delta/\kappa$, $c=|\chi_s|/\kappa$, and $u=\Omega_{\rm LT}/\kappa$. Direct inversion of Eq.~(\ref{eq:blocks}) gives
\begin{equation}
q(x,c,u)=1-\frac{c^2}{2\{[(1+c^2+u^2)/4-x^2]^2+x^2\}}.
\label{eq:q}
\end{equation}
The detunings satisfying the CZ condition obey
\begin{equation}
x_\pm^2=\frac{c^2+u^2-1}{4} \pm\frac12\sqrt{c^2-u^2}.
\label{eq:roots}
\end{equation}
A physical solution requires $c^2-u^2\ge0$ and a nonnegative value of $x_\pm^2$. The smallest interaction admitting a real detuning, together with the corresponding detuning magnitude, is
\begin{align}
c_{\min}(u)&=
\begin{cases}
\sqrt2-\sqrt{1-u^2},&u\le1/\sqrt2,\\
u,&u\ge1/\sqrt2,
\end{cases}\notag\\
|x_c|&=\begin{cases}0,&u\le1/\sqrt2,\\
\frac12\sqrt{2u^2-1},&u\ge1/\sqrt2.
\end{cases}
\label{eq:critical}
\end{align}
At $u=0$, the threshold $c_{\min}=\sqrt2-1$ is 58.6\% below the scalar-Kerr value $c=1$ for the same one-sided linewidth convention. This is a comparison of monochromatic interaction thresholds. The bandwidth and loss tradeoffs are evaluated separately in the Supplemental Material.

\begin{figure*}[t]
\centering
\includegraphics[width=.9\textwidth]{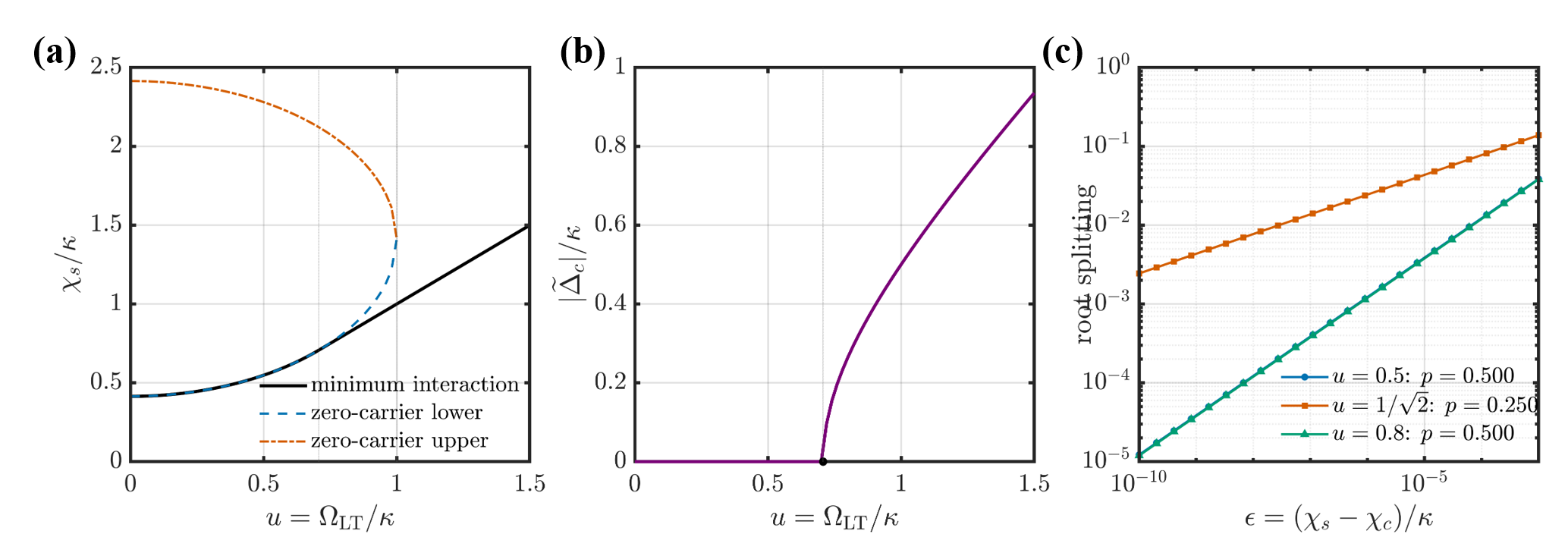}
\caption{Solutions of the CZ condition, Eq.~(\ref{eq:roots}). (a) Minimum interaction $c_{\min}(u)$ (black) and the two branches at zero detuning. The threshold changes form at $u=1/\sqrt2$, whereas the zero-detuning branches meet at $u=1$. (b) The threshold detuning becomes nonzero for $u>1/\sqrt2$. (c) Splitting away from a double root scales as $\epsilon^{1/2}$. Along a fixed-$u$ cut at $u=1/\sqrt2$, the real solutions emerge from the fourfold root as $\epsilon^{1/4}$. Lines are analytic results and markers are independent evaluations of the response matrices.}
\label{fig:topology}
\end{figure*}

\paragraph*{Solutions of the CZ condition.---}
For $u<1/\sqrt2$, the threshold solution is a double root at $x=0$, and its splitting scales as $\epsilon^{1/2}$ for $\epsilon=c-c_{\min}>0$. At $c=u=1/\sqrt2$, both branches of $x^2$ vanish and $q(x)=4x^4/(1+4x^4)$, so the CZ condition has a fourfold root. Increasing $c$ at fixed $u=1/\sqrt2$ gives real solutions proportional to $\epsilon^{1/4}$. For $u>1/\sqrt2$, the threshold instead occurs at $c=u$ and two nonzero detunings. Zero-detuning solutions also occur at the interaction strengths
\begin{equation}
c_\pm=\sqrt2\pm\sqrt{1-u^2},\qquad u\le1,
\label{eq:central}
\end{equation}
which meet at $u=1$. Figure~\ref{fig:topology} shows that transverse splitting changes both the number and positions of the CZ solutions. The degeneracy at $c=u=1/\sqrt2$ belongs to the gate condition $q=0$ and is not an exceptional point.

To resolve the parameter dependence near this point, define $\alpha=c+u-\sqrt2$ and $\beta=c-u$. The CZ condition can then be written exactly as
\begin{align}
[x^2-m(\alpha,\beta)]^2&=\frac{\beta(\sqrt2+\alpha)}4,\notag\\
m(\alpha,\beta)&=\frac{\alpha}{2\sqrt2} +\frac{\alpha^2+\beta^2}{8}.
\label{eq:normal}
\end{align}
Near the fourfold root, the parameter combinations scale as $\alpha\sim x^2$ and $\beta\sim x^4$. For paths crossing $c=u$ transversely at this point, the real detunings split with exponent $1/4$. Along $c=u$, the splitting is square-root. The exponent depends on the path through parameter space.

For the joint limit of small bandwidth and small parameter offsets, set $\alpha=As^2$, $\beta=Bs^4$, and $x=sX$. Here $s=\sigma/\kappa$, with $\sigma$ the root-mean-square width of the Gaussian intensity spectrum. The CZ condition and leading error of the reference analyzer are
\begin{align}
[X^2-A/(2\sqrt2)]^2&=B/(2\sqrt2),\notag\\
1-F_{\rm B}&=(6+8X^2)s^4+o(s^4).
\label{eq:master}
\end{align}
The scaling variables are $\alpha/s^2$ and $\beta/s^4$. Rephasing $W_0\to W_0R_z(\gamma)$ gives fixed analyzers with the same ideal action at $\omega_0$ and error coefficient $C_\gamma=(1+\sin^2\gamma)(6+8X^2)$. We use its minimum at $\gamma=0$ modulo $\pi$. With $y=(\omega-\omega_0)/\kappa$, the off-diagonal derivative in the fixed $W_0$ basis obeys $\|\partial_y V_{\rm off}\|_F/s\to8|X|$. The term $8X^2$ comes from the first-order frequency dependence of the rotation axis, while the constant 6 comes from its second-order variation at zero detuning.

For a finite pulse, the fixed input rotation
\begin{equation}
Q_\theta=I\otimes R_y(\theta),\quad \theta=+2\sqrt2s^2,\quad U_{\rm pkt}(\omega)=U_{\rm car}(\omega)Q_\theta,
\label{eq:packet-rotation}
\end{equation}
subtracts the spectral mean of the leading quadratic amplitudes for incorrect outcomes and gives
\begin{equation}
1-F_{\rm B}^{\rm pkt}=(4+8X^2)s^4+o(s^4).
\label{eq:packet-master}
\end{equation}
The subscript $\mathrm{pkt}$ denotes this bandwidth-adjusted setting. The angle is the same for every frequency component and Bell input. Equation~(\ref{eq:packet-master}) gives the error for this choice. Since $Q_\theta$ is local, the normalized POVM elements and ideal-link remote states obey
\begin{equation}
e_\mu^{\rm pkt}=Q_\theta^\dagger e_\mu^{\rm car}Q_\theta,\qquad
\rho_\mu^{\rm pkt}=Q_\theta^{\mathsf T}\rho_\mu^{\rm car}Q_\theta^*.
\label{eq:frame-covariance}
\end{equation}
Concurrence and negativity are unchanged by this transformation. The rotation improves Bell-state identification. Its effect on coherence in the measurement basis is derived in the Supplemental Material.

\paragraph*{Finite-bandwidth Bell-state discrimination.---}
For a normalized Gaussian intensity spectrum $w(\omega)$, tracing over the unresolved frequency gives the channel
\begin{equation}
\mathcal E(\rho)=\int d\omega\,w(\omega)
S(\omega)\rho S^\dagger(\omega).
\label{eq:channel}
\end{equation}
Let $M(\omega)$ be the amplitude matrix from the ordered Bell inputs to the ordered measurement outcomes after the fixed local rotations. The unitarity requirement yields the exact error probability $1-F_{\rm B}=\int d\omega\,w\|M_{\rm off}\|_F^2/4$, where $M_{\rm off}$ contains the amplitudes for incorrect outcomes.

Without transverse splitting, the interaction threshold gives a quartic bandwidth-error coefficient $9+6\sqrt2$, with crossover variable $\epsilon/s^2$ away from the double root. Zero-detuning solutions retain quartic identification errors. At minimum-interaction solutions with nonzero detuning, the error becomes quadratic because the rotation axis varies linearly with frequency. The gate condition fixes the phase at $\omega_0$, while the finite-pulse error also depends on the local rotations across the spectrum. Figure~\ref{fig:bandwidth} compares these contributions. The derivation is given in the Supplemental Material.

\begin{figure}[t]
\centering
\includegraphics[width=.9\columnwidth]{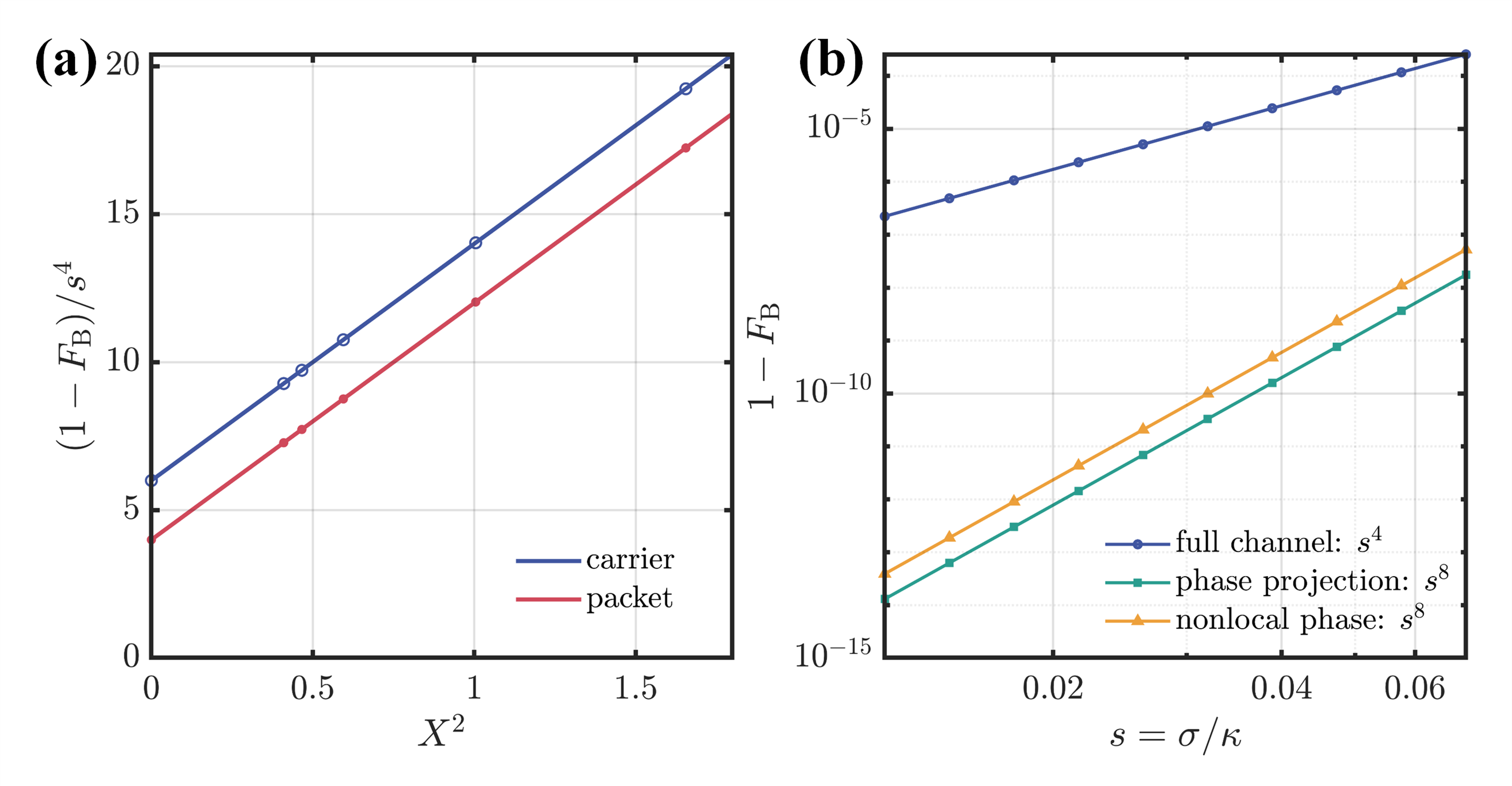}
\caption{Bell-state discrimination for finite pulses near $c=u=1/\sqrt2$, with fixed phase convention $\gamma=0$. (a) Full calculations approach error coefficients $6+8X^2$ for the reference analyzer and $4+8X^2$ with the input rotation $Q_\theta$. (b) Along $A=0,B=1$, the fixed circuit gives an error of order $s^4$. Removing the frequency dependence of the local rotations, or retaining only the nonlocal phase, gives an $s^8$ dependence. The measurement fidelities use frequency-independent local rotations.}
\label{fig:bandwidth}
\end{figure}

\paragraph*{Entanglement swapping.---}
For each accepted outcome, we evaluate the remote fidelity, concurrence, and negativity after the known Pauli correction. The average $\overline{\mathcal C}_{AB}=\sum_\mu p_\mu\mathcal C(\rho_{AB}^{(\mu)})/P_{\rm acc}$ retains the classical outcome record. It is generally different from the concurrence of the mixture obtained by discarding that record. Negativity is averaged in the same way. Figure~\ref{fig:swapping}(a) verifies $\overline F_{AB}=F_{\rm B}$ for ideal links. Loss changes the acceptance probability while preserving this identity.

For the lossless node at $c=u=1/\sqrt2$ and $x=0$, all outcomes have $p_\mu=1/4$. Figure~\ref{fig:swapping}(b) compares the remote fidelity, concurrence, and negativity calculated from the full states after tracing over frequency. Their narrow-band limits are
\begin{align}
\overline F_{AB}^{\rm car}&=1-6s^4+o(s^4),\notag\\
\overline{\mathcal C}_{AB}&=1-8s^4+o(s^4),\notag\\
2\overline{\mathcal N}_{AB}&=1-8s^4+o(s^4).
\label{eq:remote-bandwidth}
\end{align}
Remote entanglement decreases quartically with bandwidth. Dephasing the reference output in the Bell basis increases the concurrence-loss coefficient from 8 to 12. The $O(s^2)$ off-diagonal coherences contribute to the concurrence by the order $s^4$.

To separate imperfections of the sources from those of the node, take two identical Werner links $\rho_v=v|\Phi^+\rangle\langle\Phi^+|+(1-v)I_4/4$, with $0\le v\le1$. Writing $\Lambda_v(\tau)=v\tau+(1-v)\Tr(\tau)I_2/2$, the exact conditional output before the Pauli correction is
\begin{equation}
\rho_{AB}^{(\mu)}(s,v)=
(\Lambda_v\otimes\Lambda_v)\bigl[e_\mu(s)^T\bigr].
\label{eq:source-dressed-swap}
\end{equation}
For an ideal analyzer, Fig.~\ref{fig:swapping}(c) gives a remote Werner state of visibility $v^2$, with $\overline F_{AB}=(1+3v^2)/4$ and $\overline{\mathcal C}_{AB}=2\overline{\mathcal N}_{AB}=[(3v^2-1)/2]_+$, where $[z]_+=\max(0,z)$. Entanglement requires $v>1/\sqrt3$. The remote fidelity now depends on source quality as well as the measurement and differs from $F_{\rm B}$.

Figure~\ref{fig:swapping}(d) shows the joint dependence on bandwidth and link visibility. The color is $\overline{\mathcal C}_{AB}$ from the full four-qubit calculation. The solid curve marks the threshold visibility $v_c(s)$, above which the output is entangled while the dashed line is the ideal-node value $1/\sqrt3$. In the two-parameter limit of Eq.~(\ref{eq:master}), the leading Bell-basis block gives $\mathcal C_{AB}=[v^2\mathcal C_{\rm n}-(1-v^2)/2]_+$ for equal links. With $\mathcal C_{\rm n}=1-K_{\rm full}s^4+o(s^4)$ and $K_{\rm full}=8+16X^2$, the threshold is
\begin{equation}
v_c(s,X)=\frac1{\sqrt3}
+\frac{8+16X^2}{3\sqrt3}s^4+o(s^4).
\label{eq:remote-boundary}
\end{equation}
The plotted node has $X=0$. Its fourth-order bandwidth error sets the extra link visibility needed for an entangled output. Equation~(\ref{eq:remote-boundary}) gives the narrow-band limit and the plotted threshold is calculated from the complete frequency-averaged state.

Werner depolarization commutes with local unitary rotations. The added input rotation leaves the remote concurrence, negativity, and threshold visibility unchanged, while improving fidelity to the assigned Bell state. Unequal or anisotropic link noise is applied separately to each qubit. The Supplemental Material specifies where the description using only the visibility product is sufficient. The control must remain coherent throughout the target pulse. A sufficient regime is $\Gamma_c,\gamma_{\phi,c}\ll\sigma\ll\kappa$, where $\Gamma_c$ and $\gamma_{\phi,c}$ are its decay and dephasing rates. The Supplemental Material treats loss and dephasing separately. The results characterize a single conditional Bell measurement and one swapping step.

\paragraph*{Conclusion.---}
Spin-dependent polariton interactions provide a conditional polarization rotation that, together with fixed local optics, forms a Bell-state analyzer. Transverse splitting controls the available gate conditions, while the frequency dependence of the full scattering response determines the errors for finite pulses. We identify a regime with fourth-order bandwidth errors and show that an additional fixed input rotation reduces the identification error without changing the swapped entanglement. Including elementary-link noise gives the threshold visibility for an entangled remote output. The calculation reveals how the spin dynamics of the polariton Bell node leads to the high fidelity entanglement swapping.

\begin{acknowledgements}
AVK acknowledges support from Saint Petersburg State University (Research Grant No. 125022803069-4) and from the Innovation Program for Quantum Science and Technology (No. 2021ZD0302704).
\end{acknowledgements}

\bibliography{references}
\end{document}


\title{Supplemental Material for ``Polariton Bell Node for Quantum Repeaters''}

\author{Junhui Cao$^{1}$}
\author{Alexey Kavokin$^{1,2,3,4}$}
\email{a.kavokin@westlake.edu.cn}
\affiliation{$^{1}$Abrikosov Center for Theoretical Physics, Moscow Center for Advanced Studies, Moscow 141701, Russia\\
	$^{2}$School of Science, Westlake University, 18 Shilongshan Road, Hangzhou 310024, Zhejiang Province, China\\
	$^{3}$Department of Physics, St. Petersburg State University, University Embankment, 7/9, St. Petersburg, 199034, Russia\\
	$^{4}$Russian Quantum Center, Skolkovo, Moscow Region, 121205, Russia
}

\maketitle

\section{Few-particle reduction of the spin interaction}

For the circular polarizations $\pm$, the spin-conserving interaction is
\begin{align}
	H_{\rm int}={}&U_\parallel(n_{c+}n_{t+}+n_{c-}n_{t-})
	+U_\perp(n_{c+}n_{t-}+n_{c-}n_{t+}).
\end{align}
With one polariton in each spatial mode, the basis $(|++\rangle,|+-\rangle,|-+\rangle,|--\rangle)$ has interaction energies $(U_\parallel,U_\perp,U_\perp,U_\parallel)$. In this subspace,
\begin{equation}
	H_{\rm int}=U_0I+JZ_cZ_t,\qquad
	U_0=\frac{U_\parallel+U_\perp}{2},\quad
	J=\frac{U_\parallel-U_\perp}{2}.
\end{equation}
Removing the common phase leaves $U(T)=\exp(-\ii JT Z_cZ_t/\hbar)$. The diagonal phase combination invariant under local phase rotations is
\begin{equation}
	\Phi_{\rm NL}=\phi_{++}-\phi_{+-}-\phi_{-+}+\phi_{--}
	=-2(U_\parallel-U_\perp)T/\hbar.
\end{equation}
The evolution is locally equivalent to CZ when $(U_\parallel-U_\perp)T/\hbar=\pi/2$ modulo $\pi$. At phase $\pi$, it gives the nonentangling product $\operatorname{diag}(-1,1,1,-1)=-Z_cZ_t$. For $\chi_s=(U_\parallel-U_\perp)/\hbar>0$, the scattering threshold has $G=\operatorname{diag}(-\ii,1,1,-\ii)$ in our phase convention. The local rotation $P=\operatorname{diag}(1,-\ii)$ gives $(P\otimes P)G=-\ii CZ$. The negative-$\chi_s$ branch uses the conjugate convention.

The scattering calculation holds the control occupation at $n_c=1$ and includes the zero- and one-target sectors. The mean interaction shift enters the target detuning as $\widetilde\Delta=\Delta-U_0/\hbar$. We assume $[H_{\rm node},Z_c]=0$, excluding the exchange term $c_+^\dagger c_-t_-^\dagger t_++{\rm H.c.}$ and transverse control precession. The two logical modes are distinguishable, and $U_{\parallel,\perp}$ include their mode-overlap integral.

\section{One-sided reflection and the double root of the CZ condition}

The reflection amplitude of a lossless one-sided mode is
\begin{equation}
	r(x)=1-\frac{\kappa}{\kappa/2-\ii x}
	=-\frac{\kappa/2+\ii x}{\kappa/2-\ii x}.
\end{equation}
Using the mean of the interaction-shifted resonances as the detuning origin, set $d=\chi_s/2$, $r_\parallel=r(\widetilde\Delta-d)$, and $r_\perp=r(\widetilde\Delta+d)$. For $\chi_s>0$, the selected CZ-equivalent phase relation is $r_\parallel=-\ii r_\perp$. Its mismatch factorizes as
\begin{equation}
	r_\parallel+\ii r_\perp
	=-(1+\ii)\frac{\widetilde\Delta^2+(\kappa/2)^2
		-\kappa\chi_s/2-\chi_s^2/4}
	{[\kappa/2-\ii(\widetilde\Delta-\chi_s/2)]
		[\kappa/2-\ii(\widetilde\Delta+\chi_s/2)]}.
\end{equation}
The corresponding detunings are
\begin{equation}
	\widetilde\Delta_\pm=\pm\frac12
	\sqrt{\chi_s^2+2\kappa\chi_s-\kappa^2}
\end{equation}
with real solutions for $\chi_s/\kappa\ge\sqrt2-1$. At the threshold, the two roots meet at zero detuning and the first derivative of $r_\parallel+\ii r_\perp$ vanishes. The differential group delay
\begin{equation}
	\delta\tau_g=\frac{\kappa}{(\kappa/2)^2+(\widetilde\Delta-d)^2}
	-\frac{\kappa}{(\kappa/2)^2+(\widetilde\Delta+d)^2}
\end{equation}
vanishes only at zero detuning for nonzero interaction. The threshold on the selected phase branch satisfies the CZ condition with matched group delays.

\section{Conditional scattering with a transverse field and the CZ condition}

For a control spin $\eta=\pm1$, the target Hamiltonian in angular-frequency units and its reflection matrix are
\begin{equation}
	h_\eta=\frac12(\Omega_{\rm LT}\sigma_x+\eta\chi_s\sigma_z),\qquad
	R_\eta=I-\kappa[\kappa I/2-\ii(\widetilde\Delta I-h_\eta)]^{-1}.
\end{equation}
The commutator $[h_+,h_-]=\ii\Omega_{\rm LT}\chi_s\sigma_y$ is nonzero when $\Omega_{\rm LT}\chi_s\ne0$, and the two Hamiltonians cannot be diagonalized in the same basis. The scattering operator $S=|+\rangle\langle+|\otimes R_++|-\rangle\langle-|\otimes R_-$ is a controlled unitary. Write its relative unitary as
\begin{equation}
	V=R_+^\dagger R_-=W\operatorname{diag}(e^{\ii\alpha},e^{\ii\beta})W^\dagger.
\end{equation}
The fixed local transformations
\begin{align}
	U_{\rm pre}&=I\otimes W,\\
	U_{\rm post}&=\operatorname{diag}(1,e^{-\ii\alpha})
	\otimes W^\dagger R_+^\dagger
\end{align}
give
\begin{equation}
	U_{\rm post}SU_{\rm pre}
	=\operatorname{diag}(1,1,1,e^{\ii(\beta-\alpha)}).
\end{equation}
The gate is locally equivalent to CZ for $\Tr V=0$. The eigenvectors in $W$ and the local phase correction are evaluated at the pulse's central frequency and held fixed across its spectrum.

In terms of $x=\widetilde\Delta/\kappa$, $c=\chi_s/\kappa$, and $u=\Omega_{\rm LT}/\kappa$, the matrix inverse gives
\begin{equation}
	q\equiv\frac12\Tr V
	=1-\frac{c^2}{2\{[(1+c^2+u^2)/4-x^2]^2+x^2\}}.
\end{equation}
Solving $q=0$ gives
\begin{equation}
	x_\pm^2=\frac{c^2+u^2-1}{4}\pm\frac12\sqrt{c^2-u^2}.
\end{equation}
Real detunings require $c^2-u^2\ge0$ and a nonnegative $x_\pm^2$ on at least one branch. These conditions give the interaction threshold in the main text. At zero detuning, the allowed interaction strengths are $c_\pm=\sqrt2\pm\sqrt{1-u^2}$ for $u\le1$.

Near a double root, $x^2\propto\epsilon$ gives $|x|\propto\epsilon^{1/2}$. At $c=u=1/\sqrt2$, four roots coincide at zero detuning. Their separation along different parameter paths follows from
\begin{equation}
	\left[x^2-\frac{c^2+u^2-1}{4}\right]^2
	=\frac{c^2-u^2}{4}.
\end{equation}
With $\alpha=c+u-\sqrt2$ and $\beta=c-u$, this becomes
\begin{align}
	[x^2-m]^2&=\frac{\beta(\sqrt2+\alpha)}{4},\\
	m&=\frac{\alpha}{2\sqrt2}+\frac{\alpha^2+\beta^2}{8}.
\end{align}
Near this point, $\alpha$ scales as $x^2$ and $\beta$ as $x^4$. Fixed-$u$ and $\alpha=0$ cuts give $|x|\propto\epsilon^{1/4}$; the tangent path $c=u$ gives $|x|\propto\epsilon^{1/2}$. The exact roots reproduce these path-dependent exponents.

\section{Bell-state analyzer and entanglement-swapping conventions}
\label{sec:s-swapping}

At the chosen central frequency $\omega_0$, the local rotations that convert the scattering operator into a controlled-phase gate are
\begin{align}
	V_0&=R_+^\dagger R_-
	=W_0\operatorname{diag}(e^{\ii\phi_0},e^{\ii\phi_1})W_0^\dagger,\notag\\
	P_c&=\operatorname{diag}(1,e^{-\ii\phi_0}),\qquad
	L_{\rm pre}=I\otimes W_0,\notag\\
	L_{\rm post}&=P_c\otimes[W_0^\dagger R_+^\dagger(\omega_0)].
	\label{eq:s-local-cz-factors}
\end{align}
With the rightmost matrix acting first,
\begin{equation}
	L_{\rm post}S(\omega_0)L_{\rm pre}
	=\operatorname{diag}(1,1,1,e^{\ii(\phi_1-\phi_0)}),
	\label{eq:s-local-cz}
\end{equation}
which is CZ when the detuning satisfies the gate condition. Adding the Hadamard rotations gives
\begin{align}
	U_{\rm car}(\omega)&=(H\otimes H)L_{\rm post}S(\omega)L_{\rm pre}(I\otimes H),\notag\\
	U_{\rm pre}^{\rm car}&=L_{\rm pre}(I\otimes H),\qquad
	U_{\rm post}^{\rm car}=(H\otimes H)L_{\rm post}.
	\label{eq:s-full-analyzer-factors}
\end{align}
We call this the reference analyzer and retain the label $\mathrm{car}$. Its fixed local rotations implement an ideal Bell measurement at $\omega_0$. The Hadamard operations form part of the measurement circuit.

\subsection{Phase convention for the fixed analyzer rotations}
\label{sec:s-fixed-calibration}

We use $Y=\left(\begin{smallmatrix}0&-\ii\\\ii&0\end{smallmatrix}\right)$, $R_a(\vartheta)=\exp(-\ii\vartheta\sigma_a/2)$, and the real Bell states $|\Phi^\pm\rangle=(|00\rangle\pm|11\rangle)/\sqrt2$ and $|\Psi^\pm\rangle=(|01\rangle\pm|10\rangle)/\sqrt2$. For $\chi_s>0$ at $c=u=1/\sqrt2$ and $x=0$, $V_0=\ii Y$. Ordering the eigenvalues as $(-\ii,+\ii)$ fixes
\begin{equation}
	W_* = \frac1{\sqrt2}
	\begin{pmatrix}1&1\\-\ii&\ii\end{pmatrix},\qquad
	P_c=\operatorname{diag}(1,\ii).
	\label{eq:s-merger-gauge}
\end{equation}
This phase convention fixes the signs of the Bell-basis coherences and of the input rotation used below.

At a neighboring CZ solution, $V_0=\ii(n_{0y}Y+n_{0z}Z)$, where $n_{0a}=\Tr(\sigma_aV_0)/(2\ii)$ and $n_{0y}^2+n_{0z}^2=1$. A continuous choice of eigenvector phases at $\omega_0$ is
\begin{equation}
	\vartheta_0=\operatorname{atan2}(n_{0z},n_{0y}),\qquad
	W_0=R_x(\vartheta_0)W_* .
	\label{eq:s-continuous-carrier-gauge}
\end{equation}
This gives $W_0^\dagger V_0W_0=\operatorname{diag}(-\ii,+\ii)$. Rephasing the eigenvectors defines the family of fixed analyzer settings
\begin{align}
	W_{0,\gamma}&=W_0D_\gamma,\qquad
	D_\gamma=R_z(\gamma)
	=\operatorname{diag}(e^{-\ii\gamma/2},e^{\ii\gamma/2}),\notag\\
	L_{{\rm pre},\gamma}&=I\otimes W_{0,\gamma},\qquad
	L_{{\rm post},\gamma}=P_c\otimes
	[W_{0,\gamma}^\dagger R_+^\dagger(\omega_0)].
	\label{eq:s-explicit-phase-family}
\end{align}
All matrices are fixed at $\omega_0$ during the frequency integral. Main-text Figs.~3 and~4 use $\gamma=0$. Varying $\gamma$ changes the physical analyzer setting; the corresponding effect on bandwidth errors is calculated below. The negative-$\chi_s$ branch uses the conjugate phase convention.

The bandwidth-adjusted setting, labeled $\mathrm{pkt}$, includes a fixed target input rotation:
\begin{align}
	Q_\theta&=I\otimes R_y(\theta),\qquad
	R_y(\theta)=e^{-\ii\theta Y/2},\notag\\
	U_{\rm pkt}(\omega)&=U_{\rm car}(\omega)Q_\theta,\qquad
	U_{\rm pre}^{\rm pkt}=U_{\rm pre}^{\rm car}Q_\theta .
	\label{eq:s-packet-analyzer-factors}
\end{align}
At $c=u=1/\sqrt2$ and $x=0$, the leading narrow-band choice is $\theta=+2\sqrt2s^2$, with $s=\sigma/\kappa$. This rotation acts first on the target and is shared by all frequency components and all four Bell inputs.

The ideal Bell-state analyzer is
\begin{equation}
	U_{\rm BSA}=(H\otimes H)CZ(I\otimes H),
\end{equation}
which maps $(\Phi^+,\Phi^-,\Psi^+,\Psi^-)$ to the computational-basis outcomes $(00,10,01,11)$. With this row ordering, $M_{\mu\nu}(\omega)=\langle d_\mu|U(\omega)|B_\nu\rangle$ satisfies $M_{\rm car}(\omega_0)=I$ up to outcome phases. The additional input rotation reduces the leading frequency-averaged error by subtracting the mean of the amplitudes for incorrect outcomes; it need not preserve this equality at $\omega_0$. Regrouping two elementary pairs gives
\begin{equation}
	|\Phi^+\rangle_{Ac}|\Phi^+\rangle_{tB}
	=\frac12\sum_{\mu=0}^{3}|B_\mu\rangle_{ct}
	(I\otimes\sigma_\mu)|\Phi^+\rangle_{AB},
\end{equation}
up to the fixed Bell-state phase convention. Each accepted outcome specifies a Pauli correction on the remote pair.

For outcome $\mu$, let $a_\mu(\omega)=\langle\mu|U_{\rm post}S(\omega)U_{\rm pre}$ be the row vector of measurement amplitudes. Tracing over frequency gives the POVM element
\begin{equation}
	E_\mu=\int d\omega\,w(\omega)a_\mu^\dagger(\omega)a_\mu(\omega).
\end{equation}
Contracting the inner qubits of two ideal $\Phi^+$ links gives
\begin{equation}
	\widetilde\rho_{AB}^{(\mu)}=\frac14\int d\omega\,w(\omega)
	a_\mu^T(\omega)a_\mu^*(\omega)=\frac14E_\mu^T,
	\qquad p_\mu=\frac14\Tr E_\mu .
	\label{eq:s-choi-swap}
\end{equation}
The transpose is taken in the computational basis. This contraction also applies to trace-nonincreasing measurement operations. In terms of the conditional outcome probabilities $C_{\nu\mu}=p(\mu|B_\nu)$, the remote fidelity for outcome $\mu$ is
\begin{equation}
	F_{AB}^{(\mu)}=\frac{C_{\mu\mu}}{\sum_\nu C_{\nu\mu}},
\end{equation}
with the denominator given by a column sum of $C$. Correct identification conditioned on a Bell input uses a row sum, so the two fidelities can differ for asymmetric or lossy measurements. Averaging over accepted outcomes gives
\begin{equation}
	\frac{\sum_\mu p_\mu F_{AB}^{(\mu)}}{\sum_\mu p_\mu}
	=\frac{P_{\rm corr}}{P_{\rm acc}}=F_{\rm B}.
	\label{eq:s-weighted-swap}
\end{equation}
The Pauli corrections on $B$ for outcomes $(\Phi^+,\Phi^-,\Psi^+,\Psi^-)$ are $(I,Z,X,XZ)$. We retain the outcome record and average the corrected-state concurrence as $\sum_\mu p_\mu\mathcal C(\rho_\mu)/P_{\rm acc}$. Discarding the record gives the mixture with concurrence $\mathcal C(\sum_\mu p_\mu\rho_\mu/P_{\rm acc})$, which is generally different.

For two identical Werner links $\rho_v=v|\Phi^+\rangle\langle\Phi^+|+(1-v)I/4$ and an ideal analyzer, the swapped state has visibility $v^2$ and
\begin{align}
	F_{AB}&=(1+3v^2)/4,\notag\\
	\mathcal C_{AB}&=\max[0,(3v^2-1)/2],\qquad
	\mathcal N_{AB}=\max[0,(3v^2-1)/4].
	\label{eq:s-werner-swap}
\end{align}
The output is entangled for $v>1/\sqrt3$. These expressions give the ideal-analyzer curves at $s=0$ in main-text Fig.~\MainSwapFig(c). The equality between remote fidelity and $F_{\rm B}$ in Eq.~(\ref{eq:s-weighted-swap}) is restricted to ideal elementary links.

\subsection{Conditional remote states and observables in main-text Fig.~\MainSwapFig}
\label{sec:s-main-figure-four}

Main-text Fig.~\MainSwapFig(a) compares Eq.~(\ref{eq:s-weighted-swap}) with direct four-qubit calculations for ideal links, including trace-nonincreasing operations. The ratio $P_{\rm corr}/P_{\rm acc}$ retains the probability of acceptance.

For general elementary links, write $\rho_{\rm in}=\rho_{Ac}\otimes\rho_{tB}$ in the order $(A,c,t,B)$. A row Kraus operator $a_{\mu r}(\omega)$ describes accepted outcome $\mu$, with $r$ labeling an unobserved branch. The unnormalized remote state is
\begin{align}
	\widetilde\rho_{AB}^{(\mu)}
	&=\sum_r\int d\omega\,w(\omega)
	\mathcal A_{\mu r}(\omega)\rho_{\rm in}
	\mathcal A_{\mu r}^\dagger(\omega),\notag\\
	\mathcal A_{\mu r}(\omega)&=I_A\otimes a_{\mu r}(\omega)\otimes I_B,
	\qquad p_\mu=\Tr\widetilde\rho_{AB}^{(\mu)}.
	\label{eq:s-four-qubit-contraction}
\end{align}
The sum over $r$ is absent for a single Kraus operator. In general, $E_\mu=\sum_r\int d\omega\,w\,a_{\mu r}^\dagger a_{\mu r}$. For $p_\mu>0$, the Pauli correction $P_\mu=I_A\otimes\sigma_\mu^\dagger$ gives $\rho_{AB,\mu}^{\rm corr}=P_\mu\widetilde\rho_{AB}^{(\mu)}P_\mu^\dagger/p_\mu$. We plot
\begin{align}
	P_{\rm acc}&=\sum_\mu p_\mu,\notag\\
	\overline F_{AB}
	&=\frac{1}{P_{\rm acc}}\sum_\mu p_\mu
	\langle\Phi^+|\rho_{AB,\mu}^{\rm corr}|\Phi^+\rangle,\notag\\
	\overline{\mathcal C}_{AB}
	&=\frac{1}{P_{\rm acc}}\sum_\mu p_\mu
	\mathcal C(\rho_{AB,\mu}^{\rm corr}),\notag\\
	\overline{\mathcal N}_{AB}
	&=\frac{1}{P_{\rm acc}}\sum_\mu p_\mu
	\mathcal N(\rho_{AB,\mu}^{\rm corr}).
	\label{eq:s-main-figure-metrics}
\end{align}
The negativity is defined as $\mathcal N(\rho)=(\|\rho^{T_B}\|_1-1)/2$. Local Pauli corrections leave concurrence and negativity unchanged. Both are evaluated for each conditional state before averaging.

Panel (b) uses ideal links and the lossless reference analyzer at $c=u=1/\sqrt2$, $x=0$, and $\gamma=0$. The curves use the full frequency-integrated POVM elements; their narrow-band coefficients follow from Eq.~(\ref{eq:s-coherent-entanglement-law}). Panel (d) uses the same node with equal Werner links. Their maximally mixed inner-qubit marginals give $p_\mu=\Tr E_\mu/4=1/4$, independent of $v$. The color shows the average concurrence in Eq.~(\ref{eq:s-main-figure-metrics}). The solid line is the threshold visibility $v_c(s)$, and the dashed line is the ideal-node value $1/\sqrt3$. The link-noise channels and narrow-band threshold are derived in Secs.~\ref{sec:s-source-dressing} and~\ref{sec:s-survival-boundary}.

\section{Frequency averaging and Bell-state discrimination probabilities}

For an intensity spectrum normalized as $\int w(\omega)d\omega=1$,
\begin{equation}
	\mathcal E(\rho)=\int d\omega\,w(\omega)S(\omega)\rho S^\dagger(\omega).
\end{equation}
If $S(\omega)=\operatorname{diag}(s_0,\ldots,s_3)$, the map acts by elementwise multiplication, also known as a Schur product,
\begin{equation}
	[\mathcal E(\rho)]_{jl}=G_{jl}\rho_{jl},\qquad
	G_{jl}=\int d\omega\,w(\omega)s_j(\omega)s_l^*(\omega).
\end{equation}
The full matrix calculation agrees with this Gram-matrix form for diagonal $S$. For each Bell input $\nu$, the fixed analyzer gives $p_{\mu|\nu}=\langle B_\nu|E_\mu|B_\nu\rangle$. Averaging over the four equally likely inputs defines
\begin{equation}
	P_{\rm corr}=\frac14\sum_\nu p_{\nu|\nu},\quad
	P_{\rm acc}=\frac14\sum_{\mu,\nu}p_{\mu|\nu},\quad
	F_{\rm B}=P_{\rm corr}/P_{\rm acc}.
\end{equation}
Here $P_{\rm corr}$ and $P_{\rm acc}$ are the unconditional probabilities of correct identification and acceptance. Their ratio $F_{\rm B}$ is the probability of correct identification among accepted events.

\section{Bandwidth expansion near the interaction threshold}

\begin{figure}[t]
	\centering
	\includegraphics[width=.88\textwidth]{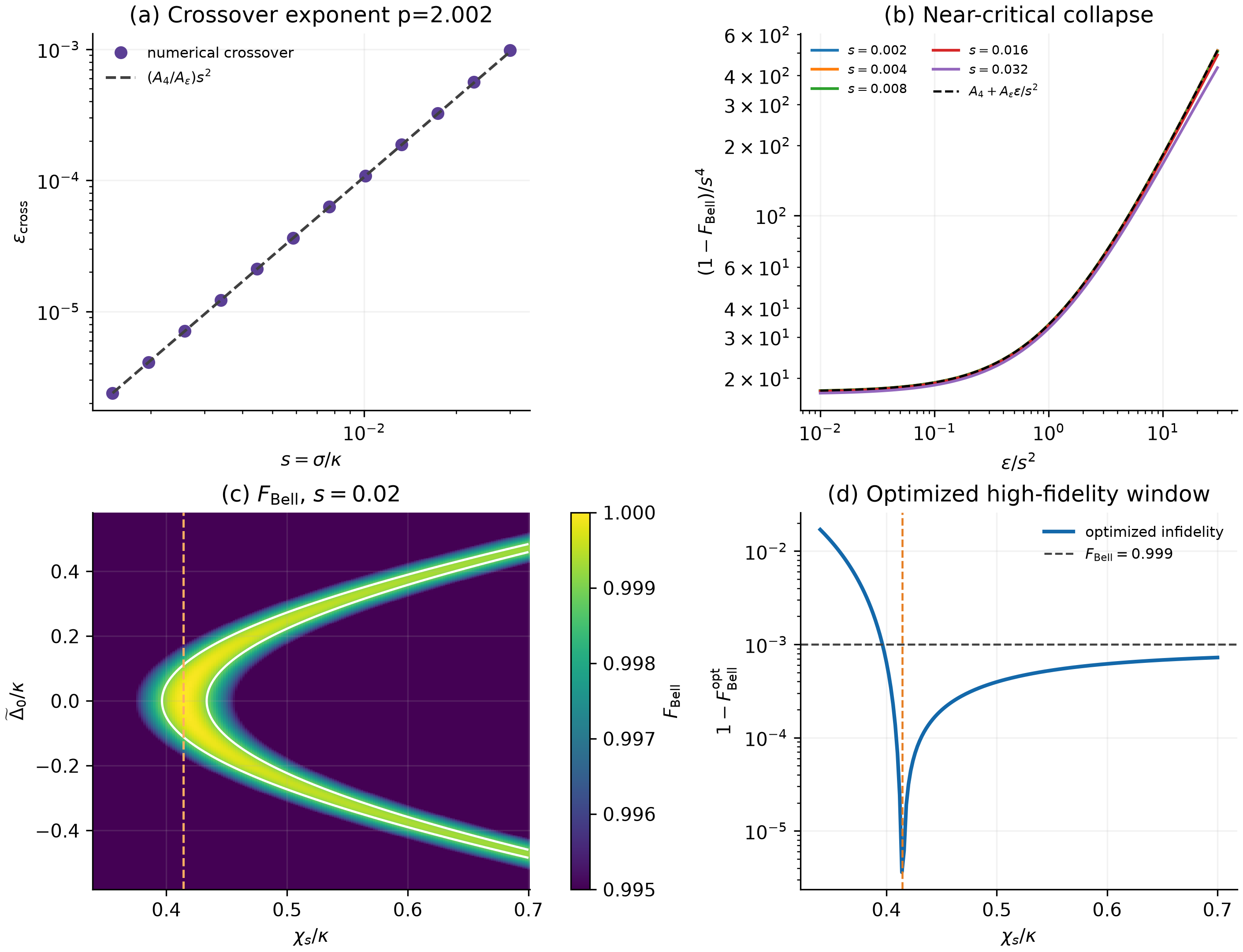}
\caption{Interaction and bandwidth dependence without TE--TM splitting. The crossover follows $\epsilon\sim s^2$, and the rescaled data agree with the analytic expansion. The lower panels show the probability of correct identification conditioned on acceptance as a function of interaction strength and detuning.}
	\label{fig:s-crossover}
\end{figure}

At the threshold of the diagonal response, fixed local phases give the identification error at frequency offset $x$ as
\begin{equation}
	p_{\rm err}(x)=\frac14|r_\parallel+\ii r_\perp|^2
	=\frac{x^4}{2[(\kappa/2)^2+(x-d)^2][(\kappa/2)^2+(x+d)^2]}.
\end{equation}
A Gaussian intensity spectrum of rms width $\sigma$ has $\langle x^4\rangle=3\sigma^4$. Averaging the error gives
\begin{equation}
	1-F_{\rm B}=(9+6\sqrt2)(\sigma/\kappa)^4+O[(\sigma/\kappa)^6].
\end{equation}
For $c=c_0+\epsilon$, set the central detuning to the positive CZ solution, $x_0^2=\epsilon/\sqrt2+\epsilon^2/4$. Expansion about $x_0$ gives
\begin{equation}
	1-F_{\rm B}=(8+6\sqrt2)\epsilon s^2
	+(9+6\sqrt2)s^4+\cdots,\qquad s=\sigma/\kappa.
\end{equation}
The two leading contributions are equal when
\begin{equation}
	\epsilon_{\rm cross}/s^2
	=\frac{9+6\sqrt2}{8+6\sqrt2}.
\end{equation}
The numerical crossover follows $\epsilon_{\rm cross}\propto s^2$, and the narrow-band data follow the rescaled expansion in Fig.~\ref{fig:s-crossover}.

For the matrix response, a CZ solution at zero detuning gives an error even in $x$ and quartic in bandwidth. At threshold solutions away from zero detuning, the rotation axis of $V(\omega)$ varies linearly near $\omega_0$, producing a quadratic bandwidth error. The calculation uses a fixed $W$ across the spectrum.

\section{Two-parameter scaling near the fourfold root}

To study the joint limit of small bandwidth and small parameter offsets, set
\begin{equation}
	\alpha=As^2,\qquad \beta=Bs^4,\qquad x=sX,\qquad s=\sigma/\kappa .
\end{equation}
Keeping $A$ and $B$ finite as $s\to0$ gives
\begin{equation}
	\left[X^2-\frac{A}{2\sqrt2}\right]^2=\frac{B}{2\sqrt2},
\end{equation}
The two squared detunings are
\begin{equation}
	X_\pm^2=\frac{A}{2\sqrt2}\pm2^{-3/4}\sqrt B
	+\mathcal O(s^2),
\end{equation}
on branches where the right-hand side is nonnegative. Expanding the scattering matrix in the fixed reference basis and averaging over the Gaussian spectrum gives
\begin{equation}
	1-F_{\rm B}=(6+8X_\pm^2)s^4+o(s^4).
	\label{eq:s-master}
\end{equation}
The scaled parameters are $A=\alpha/s^2$ and $B=\beta/s^4$. The phase $\gamma$ specifies the analyzer setting and leaves the CZ solutions unchanged.

At $c=u=1/\sqrt2$ and $x=0$, write $y=(\omega-\omega_0)/\kappa$. The reference analyzer has leading error amplitudes $M_{\rm off}^{\rm car}=\mathsf A y^2+o(s^2)$, with $\|\mathsf A\|_F^2=8$. The target-rotation generator satisfies $\mathsf A=-2\sqrt2\mathsf B$. Choosing
\begin{equation}
	\theta=2\sqrt2s^2,\qquad
	U_{\rm pre}^{\rm pkt}=U_{\rm pre}^{\rm car}[I\otimes R_y(\theta)]
	\label{eq:s-packet-rotation}
\end{equation}
gives
\begin{equation}
	M_{\rm off}^{\rm pkt}=\mathsf A(y^2-s^2)+o(s^2).
	\label{eq:s-centered-curvature}
\end{equation}
For a Gaussian spectrum, $\langle y^4\rangle=3s^4$ and $\operatorname{var}(y^2)=2s^4$. Subtracting the spectral mean changes the quadratic-amplitude contribution from $6s^4$ to $4s^4$. The linear-frequency contribution is unchanged, giving
\begin{equation}
	1-F_{\rm B}^{\rm pkt}=(4+8X^2)s^4+o(s^4).
	\label{eq:s-packet-master}
\end{equation}
This fixed rotation gives an achievable reduction of the identification error; global optimality over all fixed analyzers is not established. Its action on each normalized POVM element is a local unitary, which preserves concurrence and negativity for ideal links.

The frequency dependence can be resolved in the fixed basis $W_0$ that diagonalizes $V(\omega_0)=R_+^\dagger R_-$ at a CZ solution. With $y=(\omega-\omega_0)/\kappa$,
\begin{equation}
	\frac{\|[W_0^\dagger(\partial_y V)W_0]_{\rm off}\|_F}{s}
	\longrightarrow 8|X_\pm|.
\end{equation}
The squared derivative norm determines the $8X^2$ term. The coefficient 6 comes from the second-order frequency dependence at $X=0$. Two mathematical comparisons give errors of order $s^8$: replacing each transformed response by the phases of its diagonal entries, and retaining only the controlled phase up to local unitaries. Both modify the frequency-dependent response beyond the allowed fixed rotations.

\section{Bell-state discrimination errors for fixed analyzer settings}

Let $M(y)$ map the ordered Bell inputs to the ordered measurement outcomes, with correct outcomes on the diagonal. For the reference analyzer, $M(0)=I$ up to outcome phases. Unitarity gives
\begin{equation}
	1-F_{\rm B}=\frac14\int dy\,w(y)\|M_{\rm off}(y)\|_F^2 .
	\label{eq:s-leakage-identity}
\end{equation}
Here $M_{\rm off}$ contains the amplitudes for incorrect outcomes within the logical two-qubit space.

Expand $M_{\rm off}=y M_{1,\rm off}+y^2M_{2,\rm off}/2+\cdots$. For a centered Gaussian, odd spectral moments vanish. In the two-parameter limit, $M_{1,\rm off}=O(s)$ and the $M_1$--$M_3$ term is subleading, leaving
\begin{equation}
	\frac{1-F_{\rm B}}{s^4}\longrightarrow\frac14\left[
	\frac{\|M_{1,\rm off}\|_F^2}{s^2}
	+\frac34\|M_{2,\rm off}\|_F^2\right].
	\label{eq:s-geometric-expansion}
\end{equation}
With $A_\eta(y)=I/2-\ii[(x_0+y)I-h_\eta/\kappa]$, the derivatives are $R'_\eta=-\ii A_\eta^{-2}$ and $R''_\eta=2A_\eta^{-3}$. For $\gamma=0$, the scaled limit near the fourfold root gives
\begin{equation}
	\frac{\|M_{1,\rm off}\|_F^2}{s^2}\to32X^2,
	\qquad \|M_{2,\rm off}\|_F^2\to32.
\end{equation}
The two derivative terms contribute $8X^2$ and $3(32)/16=6$, respectively, to the coefficient in Eq.~(\ref{eq:s-master}).

The phase $\gamma$ in Eq.~(\ref{eq:s-explicit-phase-family}) is free at $\omega_0$. Keeping it in the derivative calculation gives
\begin{align}
	\frac{\|M_{1,\rm off}\|_F^2}{s^2}&\to
	32X^2(1+\sin^2\gamma),\notag\\
	\|M_{2,\rm off}\|_F^2&\to32(1+\sin^2\gamma),\notag\\
	C_\gamma(X)&=(1+\sin^2\gamma)(6+8X^2).
	\label{eq:s-phase-family}
\end{align}
Within this family, the coefficient is minimized at $\gamma=0$ modulo $\pi$ and doubles at $\gamma=\pi/2$. Changing $\gamma$ changes the local rotations relative to the fixed Bell inputs and Hadamard operations. Multiplying rows or columns of $M$ by phases is a representation change that leaves $\|M_{\rm off}\|_F$ invariant.

\begin{figure}[t]
	\centering
	\includegraphics[width=.98\textwidth]{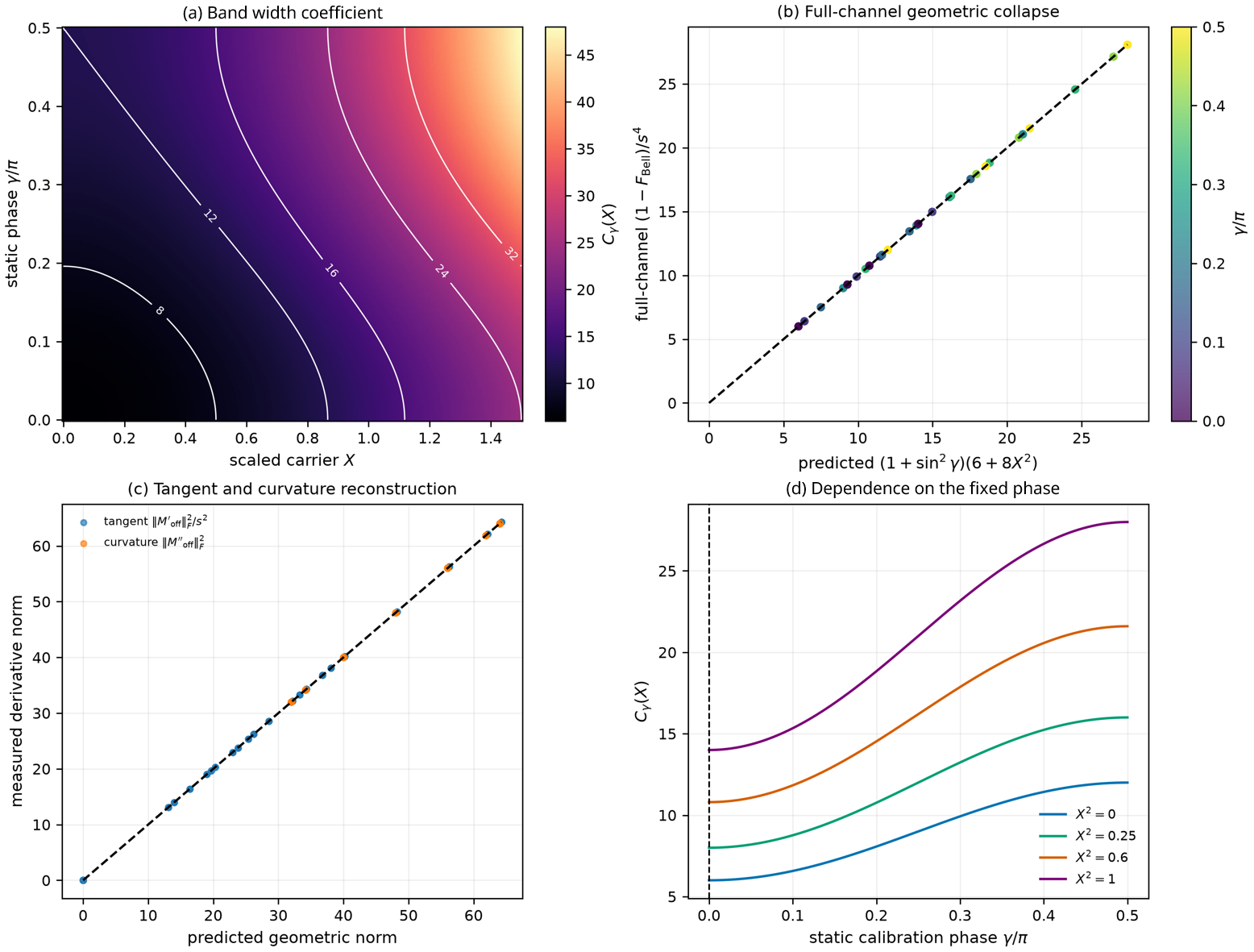}
\caption{Dependence of the bandwidth error on the analyzer setting. (a) Fourth-order coefficient versus scaled detuning and phase $\gamma$. (b) Frequency-averaged calculations and Eq.~(\ref{eq:s-phase-family}). (c) Contributions from the first and second frequency derivatives. (d) Error coefficients for analyzers with the same ideal action at $\omega_0$; $\gamma=0$ is the minimum within this family.}
	\label{fig:s-bell-geometry}
\end{figure}

\section{Measurement effects and swapped entanglement}

For ideal elementary links, Eq.~(\ref{eq:s-choi-swap}) relates the conditional remote state to the normalized POVM element $e_\mu=E_\mu/\Tr E_\mu$. For $p_\mu>0$, this gives
\begin{equation}
	\rho_{AB}^{(\mu)}=e_\mu^T,
	\qquad
	\mathcal C(\rho_{AB}^{(\mu)})=\mathcal C(e_\mu),\quad
	\mathcal N(\rho_{AB}^{(\mu)})=\mathcal N(e_\mu).
	\label{eq:s-measurement-entanglement}
\end{equation}
Full transposition complex conjugates the spectrum entering the Wootters construction and preserves the concurrence. It also preserves the partial-transpose spectrum, since $(\rho^T)^{T_B}=\rho^{T_A}=(\rho^{T_B})^T$. These normalized-state relations hold for accepted outcomes in the presence of loss.

At $c=u=1/\sqrt2$ and $x=0$, the frequency-integrated element can be expanded in $s=\sigma/\kappa$. With $\bar\mu=\mu\oplus3$ and the real Bell phases defined above,
\begin{equation}
	e_\mu=
	\begin{pmatrix}
		1-6s^4 & \xi_\mu\sqrt2s^2-6\ii s^4\\
		\xi_\mu\sqrt2s^2+6\ii s^4 & 6s^4
	\end{pmatrix}_{\{B_\mu,B_{\bar\mu}\}}
	\oplus0+O(s^6),
	\label{eq:s-effect-block}
\end{equation}
where $\xi_{0,1}=1$ and $\xi_{2,3}=-1$. These signs depend on the Bell-state phases; the coherence magnitude is phase independent. The real states $B_\mu$ and $B_{\bar\mu}$ acquire the same sign under spin flip. For a normalized block $\left(\begin{smallmatrix}a&z\\z^*&b\end{smallmatrix}\right)$ in this basis, Wootters' formula is
\begin{equation}
	\mathcal C=\sqrt{(a-b)^2+4(\Re z)^2}.
	\label{eq:s-block-concurrence}
\end{equation}
Using the expansion in Eq.~(\ref{eq:s-effect-block}),
\begin{align}
	F_{\rm B}&=1-6s^4+o(s^4),\notag\\
	\mathcal C_{\rm full}&=1-8s^4+o(s^4),\notag\\
	2\mathcal N_{\rm full}&=1-8s^4+o(s^4),\notag\\
	\mathcal C_{\rm twirl}^{\rm car}&=1-12s^4+o(s^4).
	\label{eq:s-coherent-entanglement-law}
\end{align}
We define dephasing in the fixed Bell basis by
\begin{equation}
	\mathcal T_B(\rho)=\sum_{\nu=0}^3\Pi_\nu\rho\Pi_\nu,
	\qquad \Pi_\nu=|B_\nu\rangle\langle B_\nu|.
	\label{eq:s-fixed-bell-twirl}
\end{equation}
The $O(s^2)$ coherence contributes to the concurrence at order $s^4$, accounting for the difference from the Bell-diagonal state in Fig.~\ref{fig:s-pauli-convolution}. The Bloch vector of this two-dimensional block has $r_\perp^2=4|z|^2=8s^4$ and $1-r_z=12s^4$ at leading order, with $1-r_z=(3/2)r_\perp^2+o(s^4)$.

The additional input rotation transforms the POVM elements and remote states according to
\begin{equation}
	e_\mu^{\rm pkt}=Q_\theta^\dagger e_\mu^{\rm car}Q_\theta,
	\qquad
	\rho_{AB,\mu}^{\rm pkt}=Q_\theta^{\mathsf T}
	\rho_{AB,\mu}^{\rm car}Q_\theta^* .
	\label{eq:s-frame-covariance}
\end{equation}
Both concurrence and negativity are invariant under this local rotation. At the fourfold root, the adjusted setting gives
\begin{align}
	F_{\rm B}^{\rm pkt}&=1-4s^4+o(s^4),\notag\\
	\mathcal C_{\rm full}^{\rm pkt}&=1-8s^4+o(s^4),\notag\\
	\mathcal C[\mathcal T_B(\rho^{\rm pkt})]&=1-8s^4+o(s^4).
	\label{eq:s-packet-frame-entanglement}
\end{align}
Bell-basis dephasing reduces the reference-state concurrence by $4s^4$ at leading order. This reduction vanishes at fourth order for the adjusted state. The change from 6 to 4 in the identification-error coefficient reflects an improved measurement basis, with no increase in the full-state entanglement.

\begin{figure}[t]
	\centering
	\includegraphics[width=.96\textwidth]{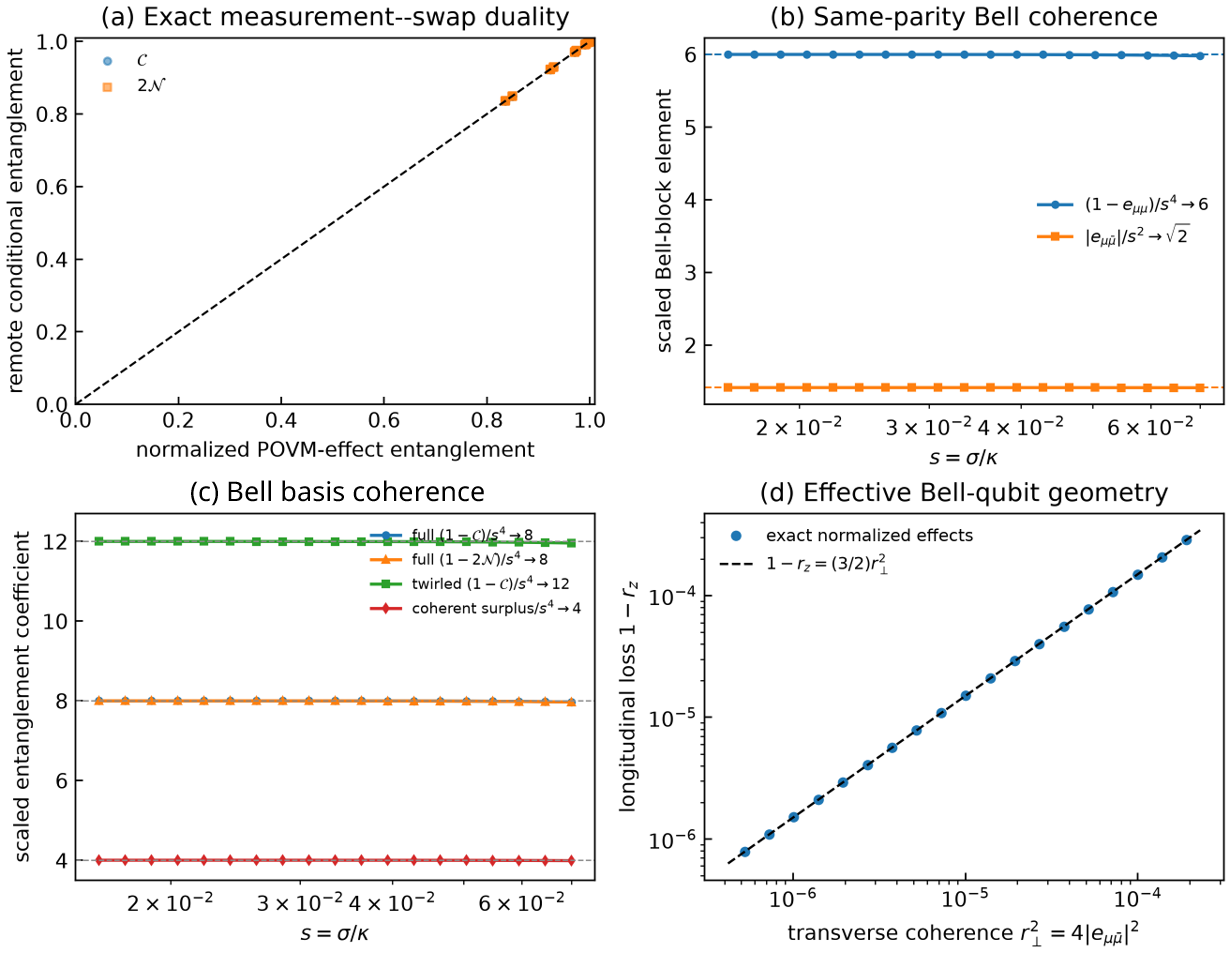}
\caption{Measurement effects and swapped entanglement. (a) Concurrence and negativity of normalized POVM elements and conditional remote states, including trace-nonincreasing operations. (b) Leading population $6s^4$ and coherence $\sqrt2s^2$ in the paired Bell-state block at $c=u=1/\sqrt2$, $x=0$. (c) Concurrence-loss coefficients of 12 for the Bell-dephased state and 8 for the full state. (d) Bloch-vector representation of the block for the reference analyzer.}
	\label{fig:s-measurement-entanglement}
\end{figure}

\subsection{Entanglement near the fourfold root}

The same Bell-state block gives the leading expansion along nearby CZ solutions. In the limit $\alpha=As^2$, $\beta=Bs^4$, $x=sX$, the allowed squared detunings are the nonnegative values of
\begin{equation}
	X_\pm^2=\frac{A}{2\sqrt2}\pm2^{-3/4}\sqrt B .
	\label{eq:s-entanglement-roots}
\end{equation}
On either branch, the frequency-integrated normalized element is
\begin{equation}
	e_\mu(X)=
	\begin{pmatrix}
		1-P_Xs^4 & \xi_\mu\sqrt2s^2+\ii O(s^4)\\
		\xi_\mu\sqrt2s^2-\ii O(s^4) & P_Xs^4
	\end{pmatrix}_{\{B_\mu,B_{\bar\mu}\}}
	\oplus0+o(s^4),\qquad P_X=6+8X^2 .
	\label{eq:s-effect-root-surface}
\end{equation}
The $8X^2$ term increases the paired-state population and leaves the leading coherence unchanged. Equation~(\ref{eq:s-block-concurrence}) gives, for each outcome,
\begin{align}
	\mathcal C_{\rm full}(X)&=1-(8+16X^2)s^4+o(s^4),\notag\\
	2\mathcal N_{\rm full}(X)&=1-(8+16X^2)s^4+o(s^4),\notag\\
	\mathcal C_{\rm twirl}^{\rm car}(X)&=1-(12+16X^2)s^4+o(s^4),\notag\\
	\mathcal C_{\rm twirl}^{\rm pkt}(X)&=1-(8+16X^2)s^4+o(s^4).
	\label{eq:s-entanglement-master}
\end{align}
For the lossless analyzer and ideal links, $\Tr E_\mu=1$ and $p_\mu=1/4$, so these expressions also describe the outcome averages. In the fixed Bell basis, the reference and adjusted settings have concurrence differences $\mathcal C_{\rm full}^{\rm car}-\mathcal C_{\rm twirl}^{\rm car}=4s^4+o(s^4)$ and $\mathcal C_{\rm full}^{\rm pkt}-\mathcal C_{\rm twirl}^{\rm pkt}=o(s^4)$. Where both solution branches exist, the full-state loss coefficients $K_\pm=8+16X_\pm^2$ differ by
\begin{equation}
	K_+-K_-=16\,2^{1/4}\sqrt B .
	\label{eq:s-entanglement-branch-split}
\end{equation}
The leading entanglement difference between branches scales as the square root of $B$. The leading coherence contribution is common to both branches.

\begin{figure}[t]
	\centering
	\includegraphics[width=.98\textwidth]{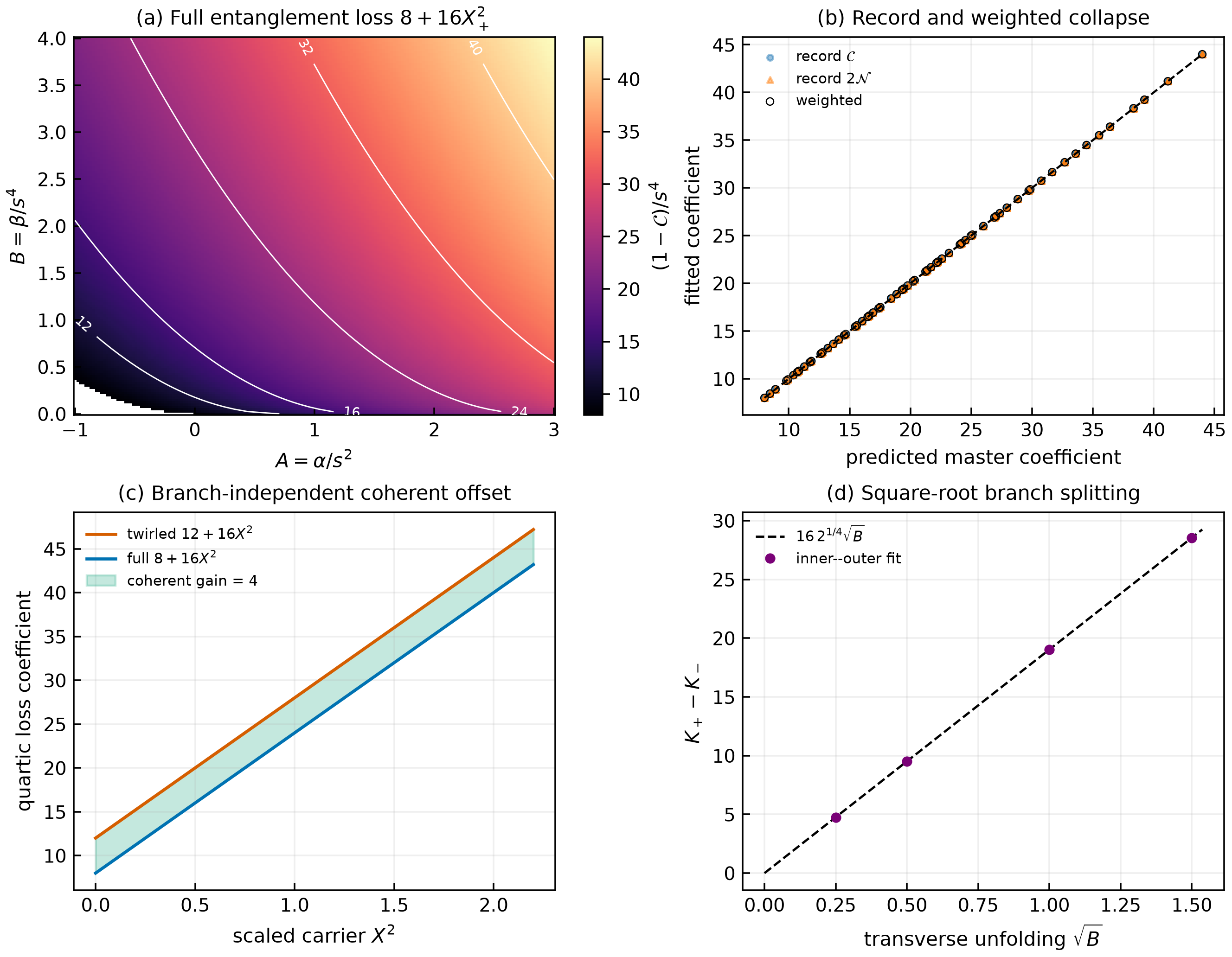}
\caption{Entanglement along nearby CZ solutions in the two-parameter limit. (a) Full-state concurrence-loss coefficient on the outer branch. (b) Coefficients from concurrence and twice-negativity fits for individual outcomes and their averages. (c) Rotation-axis dispersion raises the full-state and Bell-diagonal loss coefficients by the same amount, leaving a difference of 4 for the reference analyzer. (d) Inner--outer entanglement difference proportional to $\sqrt B$.}
	\label{fig:s-entanglement-surface}
\end{figure}

\subsection{Pauli-error composition and the limits of a fidelity-only description}

Associate the Bell states $(\Phi^+,\Phi^-,\Psi^+,\Psi^-)$ with the Pauli operators $(I,Z,X,XZ)$ modulo overall phases. Their labels form $\mathbb Z_2^2$, with bitwise XOR denoted by $\oplus$. For Bell-diagonal links with weights $p_a$ and $q_b$, the combined error distribution is
\begin{equation}
	s_e=\sum_a p_aq_{a\oplus e}.
	\label{eq:s-source-convolution}
\end{equation}
After outcome $\mu$ and its Pauli correction, the remote Bell-state populations are
\begin{equation}
	r_k^{(\mu)}=
	\frac{\sum_\nu C_{\nu\mu}s_{k\oplus\nu\oplus\mu}}
	{\sum_\nu C_{\nu\mu}}.
	\label{eq:s-record-convolution}
\end{equation}
Averaging over accepted outcomes defines the node error distribution and remote weights,
\begin{align}
	n_d&=\frac{\sum_{\nu,\mu:\,\nu\oplus\mu=d}C_{\nu\mu}}
	{\sum_{\nu,\mu}C_{\nu\mu}},\qquad F_{\rm B}=n_0,\notag\\
	r_k&=\sum_e s_e n_{k\oplus e}.
	\label{eq:s-node-convolution}
\end{align}
These populations follow from the conditional-probability matrix $C_{\nu\mu}$, with $p_\mu=\sum_\nu C_{\nu\mu}/4$ and $P_{\rm acc}=\sum_\mu p_\mu$.

For Bell-diagonal weights $r_k$, concurrence and negativity are
\begin{equation}
	\mathcal C=\max(0,2\max_k r_k-1),\qquad \mathcal N=\mathcal C/2.
	\label{eq:s-bell-diagonal-entanglement}
\end{equation}
For ideal elementary links and a Bell-diagonal remote state, specifying only $F_{\rm B}=F$ allows
\begin{equation}
	\begin{cases}
		0\le\mathcal C\le1-2F,&F\le1/2,\\
		\mathcal C=2F-1,&F\ge1/2.
	\end{cases}
	\label{eq:s-fidelity-envelope}
\end{equation}
The distributions $(0.4,0.2,0.2,0.2)$ and $(0.4,0.6,0,0)$ both have $F_{\rm B}=0.4$ but give concurrences 0 and 0.2. For two Werner links, the visibility product $V=v_Lv_R$ gives $r=Vn+(1-V)/4$. When the correctly identified sector has the largest population, the Bell-diagonal state is entangled if
\begin{equation}
	F_{\rm B}>\frac{1+V}{4V};
\end{equation}
this requires $V>1/3$.

\begin{figure}[t]
	\centering
	\includegraphics[width=.96\textwidth]{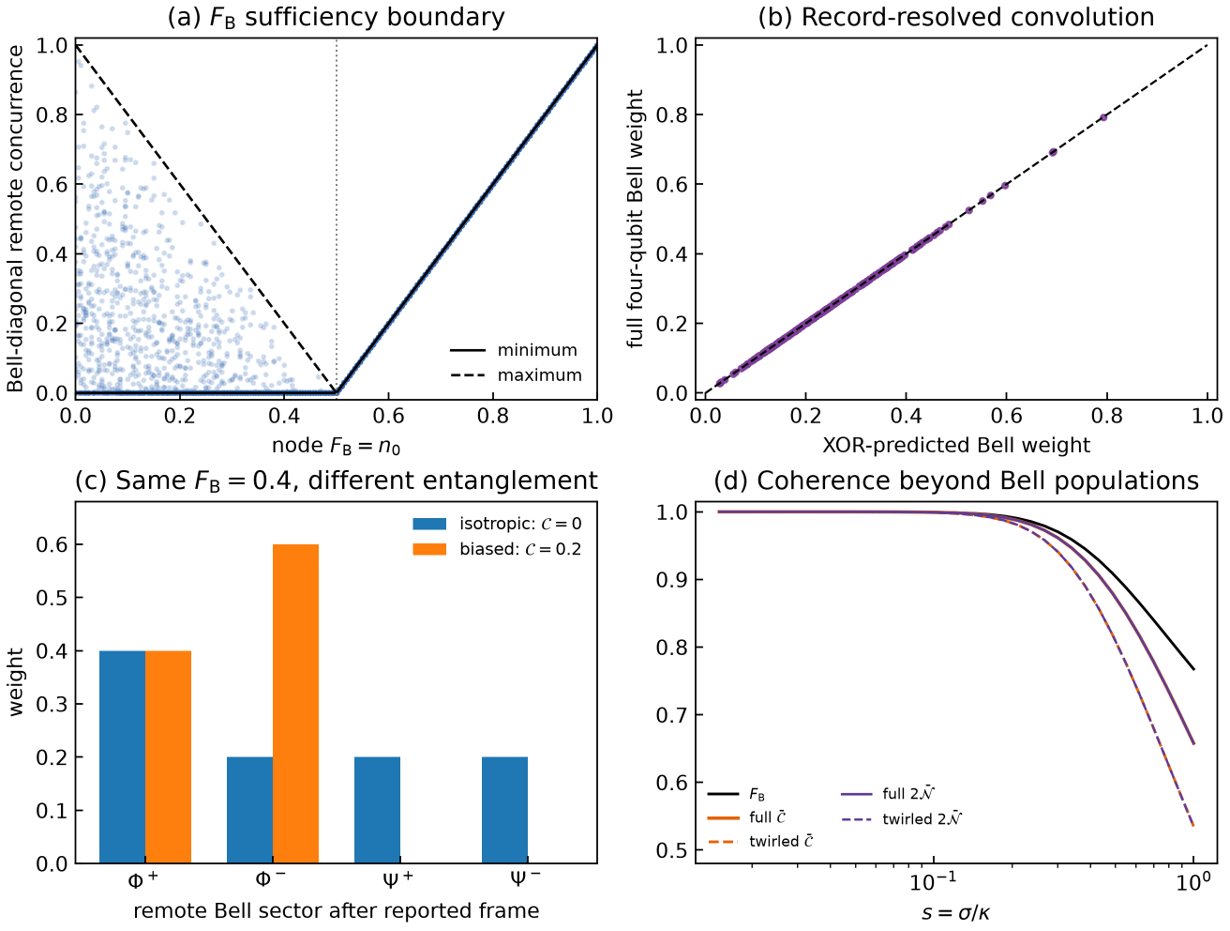}
\caption{Composition of Pauli errors. (a) Allowed concurrence at fixed $F_{\rm B}$ for ideal links and Bell-diagonal node errors. (b) Bell populations from the XOR convolution and direct four-qubit calculations. (c) Different entanglement at the same $F_{\rm B}$. (d) Full-state and Pauli-twirled entanglement for the reference analyzer.}
	\label{fig:s-pauli-convolution}
\end{figure}

The population convolution also holds for POVM elements with Bell-basis coherence. Pauli twirling removes that coherence by local operations and classical communication (LOCC); Eqs.~(\ref{eq:s-record-convolution})--(\ref{eq:s-bell-diagonal-entanglement}) then give an entanglement lower bound. Figure~\ref{fig:s-pauli-convolution} compares this bound with the full state. The reference analyzer has a fourth-order gap that vanishes for the bandwidth-adjusted setting, with dephasing defined in the same Bell basis. Calculating the full remote state requires the off-diagonal elements of $E_\mu$.

\subsection{Effects of noise in the elementary links}
\label{sec:s-source-dressing}

Bell-diagonal errors in the elementary links are represented by Pauli channels $\Lambda_L$ and $\Lambda_R$ on the outer qubits, with weights $p^L_g$ and $p^R_g$ ordered as $g=(I,Z,X,XZ)$. For $e_\mu=E_\mu/\Tr E_\mu$, the remote state before the Pauli correction is
\begin{equation}
	\rho_{AB}^{(\mu)}=(\Lambda_L\otimes\Lambda_R)(e_\mu^T).
	\label{eq:s-general-source-map}
\end{equation}
The transpose is in the computational circular-polarization basis. Each Pauli error is transferred through its own maximally entangled link, preserving local Bloch vectors and Bell-basis coherences in the calculation.

For local Bloch vectors $\bm a,\bm b$ and correlation tensor $T$, the diagonal Pauli-transfer matrices $D_L,D_R$ act as
\begin{equation}
	\bm a\mapsto D_L\bm a,\qquad \bm b\mapsto D_R\bm b,\qquad
	T_{AB}=D_L T(e_\mu^T)D_R^{\mathsf T}.
	\label{eq:s-two-sided-transfer}
\end{equation}
Reduction to a single one-qubit channel requires additional conditions on the local states and the alignment of correlation and noise axes.

The Bell populations obey $w_g=\sum_h p^L_h p^R_{h\oplus g}$. The corresponding Pauli-transfer eigenvalues are
\begin{align}
	\lambda_x&=w_I-w_Z+w_X-w_{XZ},\notag\\
	\lambda_y&=w_I-w_Z-w_X+w_{XZ},\notag\\
	\lambda_z&=w_I+w_Z-w_X-w_{XZ}.
	\label{eq:s-source-transfer}
\end{align}
For a block with maximally mixed local states and a correlation tensor aligned with the noise axes, with singular values $(1,\mathcal C_{\rm n},\mathcal C_{\rm n})$, the trace-norm criterion gives
\begin{equation}
	\mathcal C_{AB}=2\mathcal N_{AB}
	=\frac12\left[
	|\lambda_y|+\mathcal C_{\rm n}(|\lambda_x|+|\lambda_z|)-1
	\right]_+ .
	\label{eq:s-general-source-dressing}
\end{equation}
The alignment and local-state assumptions are required for Eq.~(\ref{eq:s-general-source-dressing}). As an example requiring the two-sided map, take the physical node at $c=u=1/\sqrt2$, $x=0$ with $p^L=p^R=(3/4,1/4,0,0)$. Its full POVM element gives
\begin{equation}
	\mathcal C_{AB}=\frac14-\frac{59}{10}s^4+o(s^4),
	\label{eq:s-anisotropic-example}
\end{equation}
in agreement with direct four-qubit evaluation and Eq.~(\ref{eq:s-general-source-map}) for all outcomes. Moving the combined noise to one qubit gives the incorrect coefficient 5.

For Werner links, each local noise channel is depolarizing. On an arbitrary one-qubit operator,
\begin{equation}
	\Lambda_v(X)=vX+(1-v)\Tr(X)I/2,
\end{equation}
and the joint action on $\rho_{\rm n}=e_\mu^T$, with $q=v_Lv_R$, is
\begin{multline}
	(\Lambda_{v_L}\otimes\Lambda_{v_R})(\rho_{\rm n})
	=q\rho_{\rm n}+v_L(1-v_R)\rho_A\otimes I/2\\
	{}+(1-v_L)v_R I/2\otimes\rho_B
	+(1-v_L)(1-v_R)I_4/4 .
	\label{eq:s-exact-werner-map}
\end{multline}
The local Bloch vectors depend separately on $v_L$ and $v_R$.

For a normalized block $\rho_{\rm n}|_{\{B_\mu,B_{\bar\mu}\}}=\left(\begin{smallmatrix}a&z\\z^*&b\end{smallmatrix}\right)$, with $a+b=1$ and $|z|^2\le ab$, define
\begin{equation}
	\mathcal C_{\rm n}=\sqrt{(a-b)^2+4(\Re z)^2},\qquad
	d_W=(v_L-v_R)\Im z .
\end{equation}
The spin-flip and partial-transpose spectra give
\begin{align}
	\mathcal C_{AB}&=\left[q\mathcal C_{\rm n}
	-\frac12\sqrt{(1-q)^2-4d_W^2}\right]_+,\notag\\
	2\mathcal N_{AB}&=\left[\sqrt{q^2\mathcal C_{\rm n}^2+d_W^2}
	-\frac{1-q}{2}\right]_+ .
	\label{eq:s-werner-source-dressing}
\end{align}
The two measures have the same entanglement threshold and can differ above it. For this block, $\mathcal C_{AB}=2\mathcal N_{AB}=[q\mathcal C_{\rm n}-(1-q)/2]_+$ when $v_L=v_R$, $\Im z=0$, or the local states are taken to be maximally mixed. Equation~(\ref{eq:s-effect-block}) has $\Im z=-6s^4+o(s^4)$; transposition reverses its sign. The unequal-visibility correction enters through $d_W^2=O(s^8)$. The general expressions retain the local Bloch vectors.

\subsubsection{Entanglement threshold in main-text Fig.~\MainSwapFig(d)}
\label{sec:s-survival-boundary}

Equal visibilities $v_L=v_R=v$ set $d_W=0$. For the normalized block in $\{B_\mu,B_{\bar\mu}\}$, Eq.~(\ref{eq:s-werner-source-dressing}) reduces to
\begin{equation}
	\mathcal C_{AB}=2\mathcal N_{AB}
	=\left[v^2\mathcal C_{\rm n}-\frac{1-v^2}{2}\right]_+ .
	\label{eq:s-equal-werner-block}
\end{equation}
Entanglement requires $v^2(1+2\mathcal C_{\rm n})>1$, with threshold visibility $(1+2\mathcal C_{\rm n})^{-1/2}$. Substituting $\mathcal C_{\rm n}=1-K_{\rm full}s^4+o(s^4)$ gives
\begin{equation}
	v_c=\frac1{\sqrt3}
	+\frac{K_{\rm full}}{3\sqrt3}s^4+o(s^4),\qquad
	K_{\rm full}=8+16X^2 .
	\label{eq:s-source-threshold-shift}
\end{equation}
Main-text Fig.~\MainSwapFig(d) uses $X=0$, for which the leading threshold shift is $8s^4/(3\sqrt3)$. Near the threshold,
\begin{equation}
	\mathcal C_{AB}
	=\left[\frac{3v^2-1}{2}-v^2K_{\rm full}s^4+o(s^4)\right]_+ .
	\label{eq:s-boundary-local-expansion}
\end{equation}
The positive part enforces zero concurrence in the separable region. The bandwidth correction raises the ideal threshold $1/\sqrt3$. Here $v$ is the visibility of each elementary link, whose Bell-state fidelity is $(1+3v)/4$.

Equations~(\ref{eq:s-equal-werner-block})--(\ref{eq:s-boundary-local-expansion}) describe the two-dimensional block and its narrow-band limit. Main-text Fig.~\MainSwapFig(d) uses the full frequency integral and the link-noise map in Eq.~(\ref{eq:s-general-source-map}), or equivalently Eq.~(\ref{eq:s-four-qubit-contraction}), with observables defined by Eq.~(\ref{eq:s-main-figure-metrics}). At each $s$, the threshold is the zero crossing of the smallest eigenvalue of the partially transposed remote state. All four outcomes have the same threshold for equal Werner links. Retaining the Bell-basis coherences is necessary to obtain the fourth-order shift; using $2F_{\rm B}-1$ would omit their contribution.

For equal Werner links, $\Lambda_v(U\tau U^\dagger)=U\Lambda_v(\tau)U^\dagger$ preserves the local-unitary relation between analyzer settings. Before the Pauli correction,
\begin{equation}
	\rho_{AB,\mu}^{\rm pkt}(s,v)
	=Q_\theta^{\mathsf T}\rho_{AB,\mu}^{\rm car}(s,v)Q_\theta^* .
	\label{eq:s-source-frame-covariance}
\end{equation}
Both settings have the same conditional entanglement and threshold visibility. A general anisotropic Pauli channel need not commute with $Q_\theta$, and its action must be evaluated on each qubit.

\subsubsection{Bell-basis coherence and the difference between solution branches}

Bell-basis dephasing changes $K_{\rm full}$ to $K_{\rm twirl}=K_{\rm full}+4$. In the interval $q_c^{\rm full}<q\leq q_c^{\rm twirl}$, the full state remains entangled while the dephased state is separable. On the entangled side, away from either threshold, the coherence contribution and the branch difference are
\begin{align}
	\mathcal C_{AB}^{\rm full}-\mathcal C_{AB}^{\rm twirl}
	&=4v_Lv_Rs^4+o(s^4),\notag\\
	\mathcal C_{AB,-}-\mathcal C_{AB,+}
	&=16\,2^{1/4}v_Lv_R\sqrt B\,s^4+o(s^4).
	\label{eq:s-source-dressed-gain}
\end{align}
Both fourth-order differences carry the factor $q=v_Lv_R$ when the compared states remain entangled. Unequal Werner links are described by Eq.~(\ref{eq:s-werner-source-dressing}); general anisotropic noise requires the two-sided map.

\begin{figure}[t]
	\centering
	\includegraphics[width=.98\textwidth]{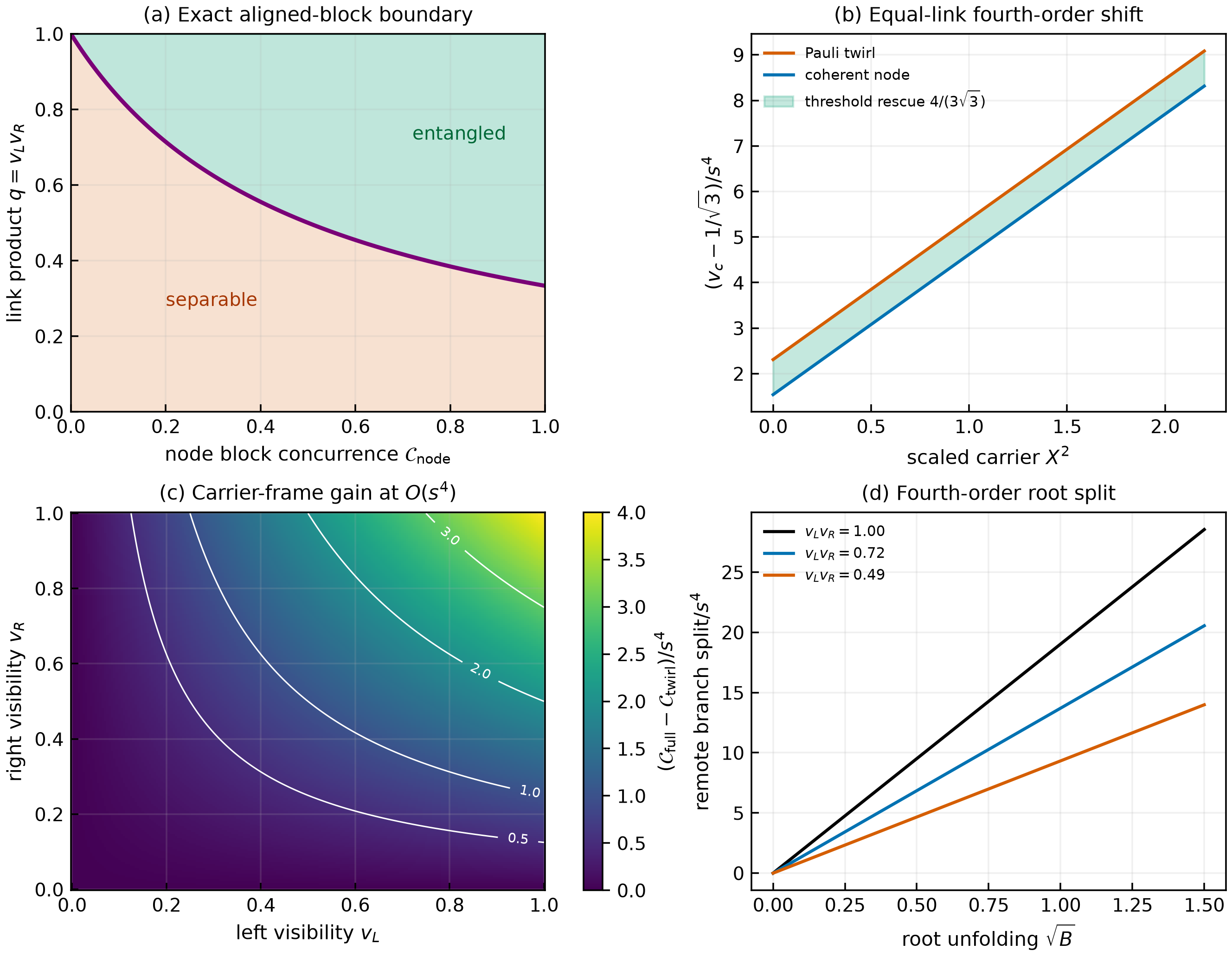}
\caption{Swapped entanglement with Werner-link noise. (a) Threshold for the aligned block, also applicable to the physical Bell block for equal visibilities or real coherence. (b) Fourth-order threshold shift; the shaded interval contains entangled full states with separable Bell-diagonal approximations for the reference analyzer. (c) Coherence contribution to concurrence, scaled by $v_Lv_R$. (d) The same factor multiplies the difference between solution branches, preserving its $\sqrt B$ dependence.}
	\label{fig:s-source-dressing}
\end{figure}

\section{Loss and imperfections of the control qubit}

The phenomenological loss calculation uses the one-sided reflection amplitude
\begin{equation}
	r_j(\Delta)=1-\frac{\kappa_{\rm ex}}
	{(\kappa_{\rm ex}+\kappa_{\rm in}+\gamma_j)/2-\ii\Delta_j}.
\end{equation}
Common internal loss and equal additional linewidths enter the denominator identically, although their microscopic origins may differ. Spin-dependent loss makes the branches unequal. At each loss value, we reoptimize the central detuning and interaction and calculate acceptance and correct-identification probabilities.

\begin{figure}[t]
	\centering
	\includegraphics[width=.88\textwidth]{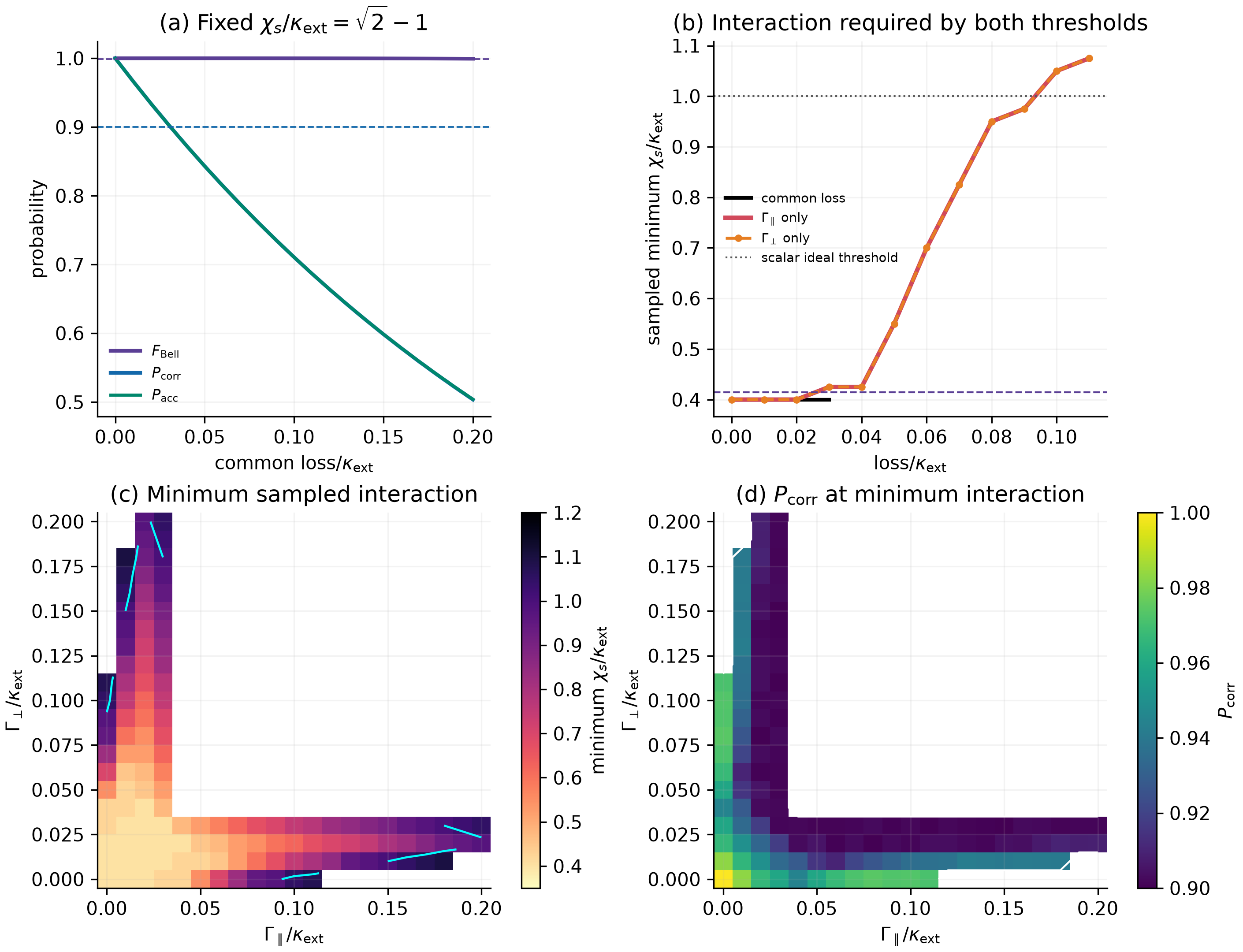}
\caption{Effects of phenomenological loss. Conditional identification fidelity, unconditional correct-outcome probability, and acceptance probability are shown separately. The calculation uses dimensionless loss parameters.}
	\label{fig:s-loss}
\end{figure}

Control survival, spin-branch imbalance, residual phase, and dephasing are described by
\begin{align}
	D&=\operatorname{diag}(\sqrt{\eta_+},\sqrt{\eta_-}e^{\ii\phi}),\\
	K_0&=\sqrt{(1+\lambda)/2}\,D,\qquad
	K_1=\sqrt{(1-\lambda)/2}\,Z D.
\end{align}
For $0\le\eta_\pm\le1$ and $|\lambda|\le1$, the operation is completely positive and trace nonincreasing. Setting $\bar\eta=(\eta_++\eta_-)/2$ and $\delta_\eta=(\eta_+-\eta_-)/(\eta_++\eta_-)$ gives, for an otherwise ideal analyzer,
\begin{equation}
	P_{\rm acc}=\bar\eta,\qquad
	F_{\rm B}=\frac{1+\lambda\sqrt{1-\delta_\eta^2}\cos\phi}{2},\qquad
	P_{\rm corr}=P_{\rm acc}F_{\rm B}.
\end{equation}
For small errors,
\begin{equation}
	1-F_{\rm B}=\frac{1-\lambda}{2}
	+\frac{\delta_\eta^2}{4}+\frac{\phi^2}{4}+\cdots.
\end{equation}
Near the fourfold root, compare control and bandwidth errors by setting
\begin{equation}
	1-\lambda=Ls^4,\qquad \delta_\eta=Ds^2,\qquad \phi=Ps^2.
\end{equation}
Composition with the frequency-averaged measurement gives
\begin{equation}
	1-F_{\rm B}=\left[C_\gamma(X)+\frac L2+\frac{D^2+P^2}{4}\right]s^4+o(s^4).
	\label{eq:s-control-master}
\end{equation}
For the reference setting $\gamma=0$, $C_\gamma=6+8X^2$. Dephasing with $1-\lambda\sim s^p$ gives a leading error exponent $\min(p,4)$. Imbalance or phase errors with $\delta_\eta\sim s^p$ or $\phi\sim s^p$ give $\min(2p,4)$. They enter at the same order as the bandwidth error for $1-\lambda=O(s^4)$ and $\delta_\eta,\phi=O(s^2)$. At finite $s$, fitted exponents vary smoothly across these asymptotic crossovers.

For equal branch survival, $\bar\eta=1-Es^4$ leaves the conditional-error coefficient $C_{\rm eff}$ in Eq.~(\ref{eq:s-control-master}) unchanged. The coefficients of $1-P_{\rm acc}$ and $1-P_{\rm corr}$ are $E$ and $C_{\rm eff}+E$. Direct POVM evaluation reproduces the ideal-analyzer expression. Finite-bandwidth results use the composed measurement and control operations.

\begin{figure}[t]
	\centering
	\includegraphics[width=.98\textwidth]{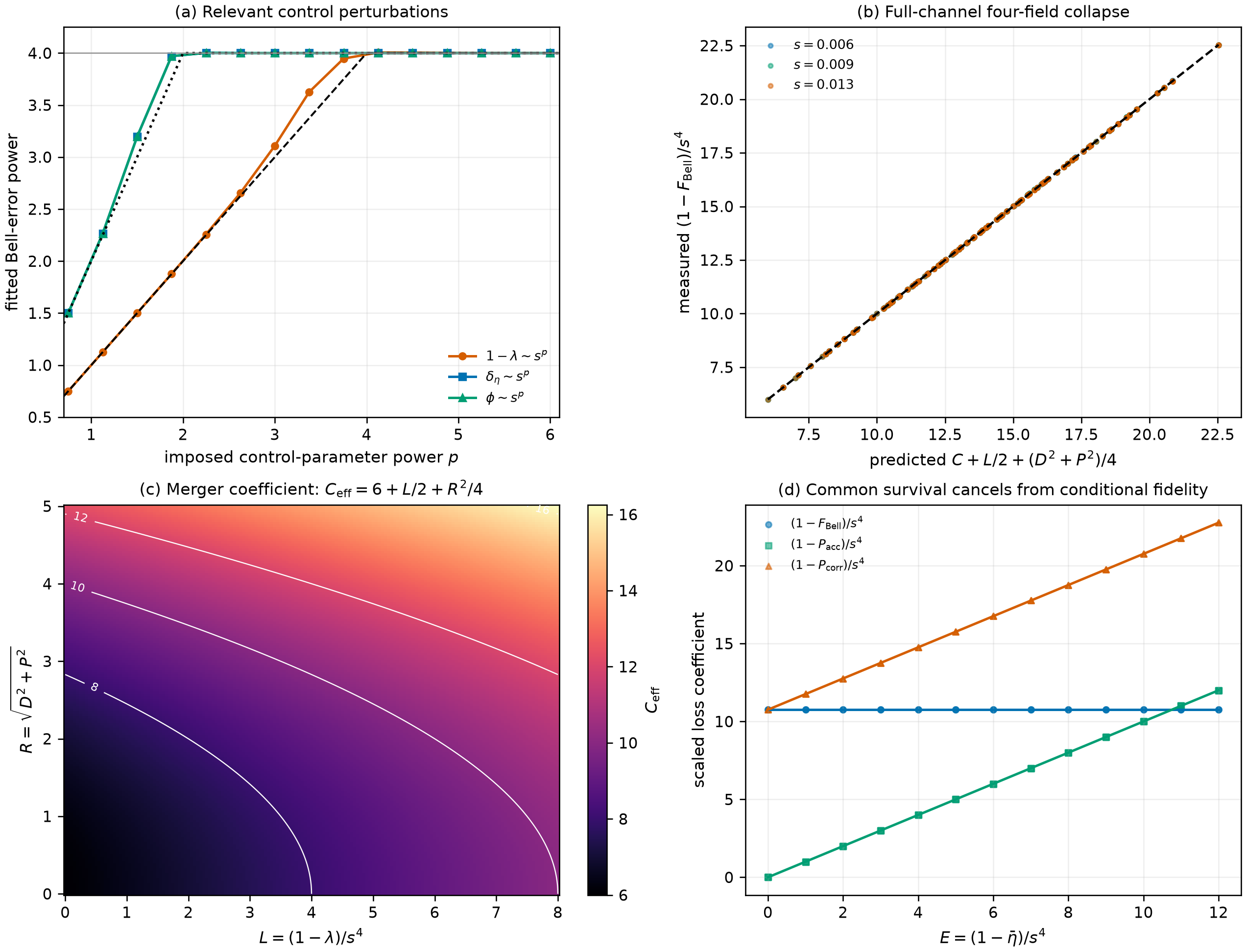}
\caption{Combined control and bandwidth errors. (a) Leading exponents and finite-$s$ crossovers. (b) Numerical coefficients and Eq.~(\ref{eq:s-control-master}). (c) Quadratic contributions from phase and survival imbalance. (d) Equal branch survival affects the event probabilities and leaves conditional fidelity unchanged. All control parameters are dimensionless.}
	\label{fig:s-control-scaling}
\end{figure}

\begin{figure}[t]
	\centering
	\includegraphics[width=.92\textwidth]{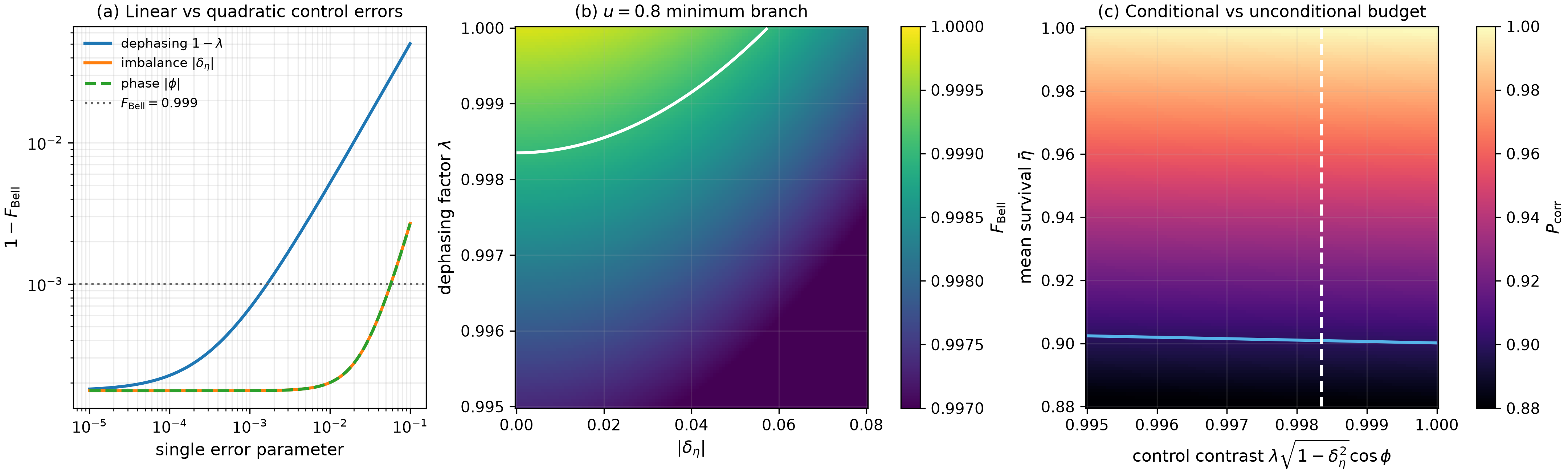}
\caption{Individual control errors. The identification error is linear in $1-\lambda$ and quadratic in small survival imbalance or phase error. The lines $F_{\rm B}=0.999$ and $P_{\rm corr}=0.90$ are reference values for conditional and unconditional performance.}
	\label{fig:s-control}
\end{figure}

\section{Comparison with a scalar Kerr gate}

The scalar reference is $S_{\rm K}=\operatorname{diag}(r_0,r_0,r_0,r_1)$, with accepted-port amplitude $r_0$ for the first three logical states. Its CZ solutions are
\begin{equation}
	\Delta_\pm=\frac{\chi\pm\sqrt{\chi^2-\kappa^2}}{2},\qquad |\chi|\ge\kappa.
\end{equation}
We compare the gates using the same linewidth, Gaussian spectral width, fixed Bell-state measurement scheme, and ideal detectors. The scalar threshold is $\chi/\kappa=1$, with quartic error coefficient 9. For the spin-dependent gate without transverse splitting, these values are $\sqrt2-1$ and $9+6\sqrt2$: a 58.6\% lower interaction threshold and a 1.943-times larger quartic coefficient on the lower branch. The minimum spin-dependent interaction remains below the scalar threshold for $u<1$ and equals it at $u=1$.

For common internal loss $l=\kappa_{\rm in}/\kappa_{\rm ex}<1$, the phase-matched points at the centers of the resonance pairs satisfy
\begin{align}
	\chi_{\rm K}/\kappa_{\rm ex}&=\sqrt{1-l^2},
	&P_{\rm K}^{(0)}&=(1-l)/(1+l),\\
	\chi_{\rm SL}/\kappa_{\rm ex}&=\sqrt{2-l^2}-1,
	&P_{\rm SL}^{(0)}&=
	\frac{[(l-1)/2]^2+[\chi_{\rm SL}/(2\kappa_{\rm ex})]^2}
	{[(l+1)/2]^2+[\chi_{\rm SL}/(2\kappa_{\rm ex})]^2}.
\end{align}
At equal common loss, the scalar gate has higher accepted-port survival, and the spin-dependent gate requires a smaller conditional splitting.

\begin{figure}[t]
	\centering
	\includegraphics[width=.92\textwidth]{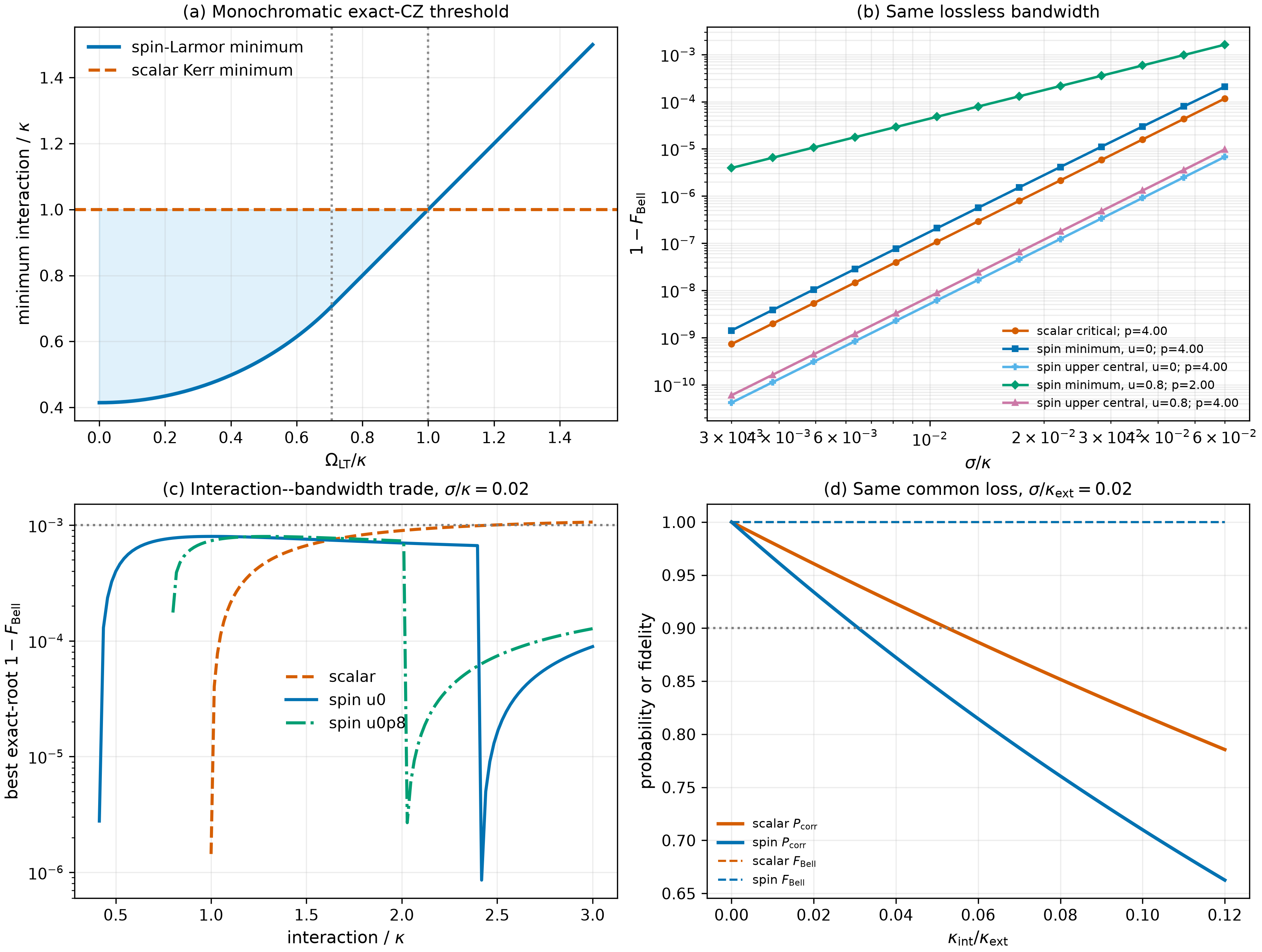}
\caption{Scalar and spin-dependent gates under the same assumptions. The panels compare interaction thresholds, spectral errors, interaction--bandwidth tradeoffs along CZ solutions, and accepted-port survival with common internal loss.}
	\label{fig:s-comparison}
\end{figure}